\documentclass[a4paper,11pt]{article}
\usepackage{jheppub} 
\usepackage{lineno}

\usepackage[utf8]{inputenc}
\usepackage[normalem]{ulem}
\usepackage{graphicx}
\usepackage{dcolumn}
\usepackage{array} 
\usepackage{bm}
\usepackage{color}
\usepackage[dvipsnames]{xcolor}

\usepackage{url}
\usepackage{xspace}
\usepackage{slashed}
\usepackage{multirow,bigstrut}
\usepackage{mathrsfs} 
\usepackage{relsize,amsmath}
\usepackage[geometry]{ifsym}
\usepackage{amssymb}
\usepackage{pifont}
\usepackage{physics}
\usepackage[compat=1.1.0]{tikz-feynman}
\usepackage{cleveref}

\usepackage{rotating}
\usepackage{ragged2e}

\usepackage{fontawesome} 
\definecolor{blue-violet}{rgb}{0.33, 0.17, 0.89}

\renewcommand{\phi}{\varphi}

\newcounter{CommentCount}
\definecolor{MH}{rgb}{0.0,0.6,9}
\definecolor{palatinate}{rgb}{0.494, 0.192, 0.482}

\definecolor{teal}{HTML}{008080}

\renewcommand{\arraystretch}{1.3}
\usepackage{siunitx}

\DeclareSIUnit \s {\second}
\DeclareSIUnit \ns {\nano\second}
\DeclareSIUnit \mus {\micro\second}
\DeclareSIUnit \ms {\milli\second}
\DeclareSIUnit \MB {\mega\byte}
\DeclareSIUnit \GB {\giga\byte}
\DeclareSIUnit \TB {\tera\byte}
\DeclareSIUnit \PB {\peta\byte}
\DeclareSIUnit \Mbps {\mega\bit/\s}
\DeclareSIUnit \Gbps {\giga\bit/\s}
\DeclareSIUnit \Tbps {\tera\bit/\s}
\DeclareSIUnit \Pbps {\peta\bit/\s}
\DeclareSIUnit \kton {\kilo\tonne} 
\DeclareSIUnit \kt {\kilo\tonne}
\DeclareSIUnit \Mt {\mega\tonne}
\DeclareSIUnit \eV {\electronvolt}
\DeclareSIUnit \keV {\kilo\electronvolt}
\DeclareSIUnit \MeV {\mega\electronvolt}
\DeclareSIUnit \GeV {\giga\electronvolt}
\DeclareSIUnit \TeV {\tera\electronvolt}
\DeclareSIUnit \PeV {\peta\electronvolt}
\DeclareSIUnit \EeV {\exa\electronvolt}
\DeclareSIUnit \m {\meter}
\DeclareSIUnit \cm {\centi\meter}
\DeclareSIUnit \in {\inchcommand}
\DeclareSIUnit \km {\kilo\meter}
\DeclareSIUnit \kV {\kilo\volt}
\DeclareSIUnit \kW {\kilo\watt}
\DeclareSIUnit \MW {\mega\watt}
\DeclareSIUnit \MHz {\mega\hertz}
\DeclareSIUnit \mrad {\milli\radian}
\DeclareSIUnit \year {years}
\DeclareSIUnit \POT {POT}
\DeclareSIUnit \sig {$\sigma$}
\DeclareSIUnit\parsec{pc}
\DeclareSIUnit\lightyear{ly}
\DeclareSIUnit\foot{ft}
\DeclareSIUnit\ft{ft}
\DeclareSIUnit \ppb{ppb}
\DeclareSIUnit \ppt{ppt}
\DeclareSIUnit \samples{S}
\DeclareSIUnit \pe{PE}
\DeclareSIUnit \T{T}

\newcommand{\enu}{\E_\enu}

\newcommand\response[1]{{\color{black} {#1}}}

\definecolor{myred}{cmyk}{0,1,1,0.55}
\definecolor{mygreen}{rgb}{0.27, 0.64, 0.48}
\definecolor{mygray}{gray}{.95}

\arxivnumber{2410.13908} 

\title{\boldmath Dark Matter Interactions in White Dwarfs: A Multi-Energy Approach to Capture Mechanisms}

\author[a]{Jaime Hoefken Zink,}

\author[b]{Shihwen Hor,}

\author[c,d]{and Maura E. Ramirez-Quezada}

\affiliation[a]{National Centre for Nuclear Research, ul. Pasteura 7, 02-093 Warsaw, Poland}
\affiliation[b]{Department of Physics, University of Tokyo, Bunkyo-ku, Tokyo 113--0033, Japan,}
\affiliation[c]{Johannes Gutenberg-Universität Mainz, 55099 Mainz, Germany, and}
\affiliation[d]{Dual CP Institute of High Energy Physics, C.P. 28045, Colima, M\'exico.}

\emailAdd{jaime.hoefkenzink@ncbj.gov.pl}
\emailAdd{shihwen@hep-th.phys.s.u-tokyo.ac.jp}
\emailAdd{mramirez@uni-mainz.de}

\abstract{White dwarfs offer a compelling avenue for probing interactions of dark matter particles, particularly in the challenging sub-GeV mass regime. The constraints derived from these celestial objects strongly depend on the existence of high dark matter densities in the corresponding regions of the Universe, where white dwarfs are observed. This implies that excluding the parameter space using local white dwarfs would present a significant challenge, primarily due to the low dark matter density in the solar neighbourhood. This limitation prompts the exploration of alternative scenarios involving dark matter particles with a diverse spectrum of kinetic energies. 
In this work, we investigate how these dark matter particles traverse the star, interact with stellar matter, and ultimately get captured.
To accomplish this, we approximate the  dark matter flux as a delta function and we assume that fermionic dark matter interacts with stellar matter either through a vector or a scalar interaction.
In our computations, we consider how interactions might vary across different energy regimes, from high-energy deep inelastic scattering and inelastic scatterings via the production of $N-$ and $\Delta-$ resonances to lower-energy elastic interactions with nucleons and nuclei. Our study models these inelastic resonant interactions with dark matter and vector or scalar mediators for the very first time. We provide insights into the specific conditions required for successfully boosted dark matter capture in white dwarfs. We found that, in general, dark matter capture is most likely to occur at low energies, as expected. However, in the high-energy regime, there remains a small window for capture through resonant and deep inelastic scattering processes.}

\begin{document}
\hfill {\tt MITP-24-075}
\maketitle
\flushbottom


\section{Introduction}

Dark matter (DM) is one of the most enigmatic components of the universe. Although it constitutes a significant portion of matter, it remains challenging to observe due to its lack of interaction with photons or other Standard Model (SM) particles through any known forces. However, it is generally assumed that DM may interact with SM particles through new physics beyond the Standard Model (BSM). To explore this possibility, various methods—direct detection, indirect detection, and collider experiments—have been pursued over the past few decades. Despite these extensive efforts, identifying DM particles remains a significant challenge in particle physics.


In general, DM particles can be captured by stars~\cite{Press:1985ug, Gould:1987ir, Gould:1987ju}. In compact stars, such as white dwarfs (WDs), their high densities enable DM particles to fall into the gravitational potential and interact much more with stellar matter, offering a promising avenue for probing DM interactions as a complementary approach to traditional detection experiments~\cite{Bertone:2007ae,McCullough:2010ai, Hooper:2010es, Amaro-Seoane:2015uny, Panotopoulos:2020kuo, Biswas:2022cyh}.
When DM particles lose sufficient energy upon scattering, they can be captured, accumulate in the WD core, undergo thermalization and eventually annihilate into SM particles. These processes could produce observable effects, allowing constraints on DM interaction strength to be established~\cite{Dasgupta:2019juq, Dasgupta:2020dik}. Although these capture mechanisms have been studied, previous treatments of the cross sections often relied on approximations that may not hold across a wide range of energies, or they have focused on specific energy ranges, such as non-relativistic processes~\cite{Bell:2021fye}. A more precise approach, accounting for different energy ranges, is needed to fully understand how sensitive compact objects like WDs are to nearby DM populations and fluxes.

Excluding DM parameter space using local WDs is challenging due to the low DM density in the solar neighborhood.\footnote{For asymmetric DM, local WD observations can feasibly exclude parameter space, as accumulating DM may collapse and trigger fusion reactions, potentially leading to a Type Ia supernova~\cite{Acevedo:2019gre}.} As a result, recent studies have focused on WDs in DM-rich environments, such as the globular cluster Messier 4 (M4)~\cite{Dasgupta:2019juq, Bell:2021fye}. In these WDs, DM-nucleon scattering can probe the sub-GeV mass range—beyond the reach of direct detection and potentially competing with conventional methods. However, these constraints rely on the presence of DM in M4. Currently, there is no evidence that globular clusters like M4 are surrounded by DM halos. These clusters may have originated from small satellite galaxies and lost much of their initial DM content through tidal stripping by the host galaxy~\cite{Bromm:2002jt, Mashchenko:2004hk, Saitoh:2005tt}. Therefore, it is important to explore other potential sources of DM flux traversing through WDs and the possibility of capture.

Recently, scenarios involving “boosted” DM (BDM) from various sources have emerged, enabling nuclear recoil signals even for lighter DM particles in direct detection experiments (e.g., see \cite{Agashe:2014yua, Wang:2021jic,Alvey:2022pad,Guha:2024mjr, Das:2024ghw}). This concept of BDM can also extend to DM capture by WDs. However, when considering accelerated DM particles colliding with a WD, the capture rate computation changes significantly. In the case of BDM, particle energies can become so high that capture may seem to be unlikely or even impossible. Nonetheless, at these higher energies, the probability of interaction increases substantially, implying that the scattering cross-section of DM could be several orders of magnitude larger than in scenarios with non-accelerated DM flux.


In this work, we present the first attempt to describe the capture of DM particles in WDs across a full energy regime, from very low energies to higher energies. This approach differs from conventional DM capture in two main ways. First, the flux of incoming energetic DM is influenced by its source, resulting in variations in the characteristics of the flux. We assume that DM is isotropically boosted from various sources throughout the universe, resulting in a flux that arrives at the star from all directions. This simplification allows us to ignore the detailed angle of impact and area of incidence. Furthermore, to account for any type of flux, we \response{fix the incoming velocity far from the star, in order to analyse the effects of particular incoming energies in the capture and see how sensitive is the star to DM at those energies.} In this way, the results for any flux can, in principle, be reconstructed from our findings, allowing us to describe a broad spectrum: from non-relativistic contributions, such as the local DM population typically modelled by a Boltzmann distribution, to BDM from other sources. 
In regions with low local DM densities, such as those around nearby WDs, traditional non-relativistic DM populations are insufficient to enable significant capture. BDM helps to overcome this limitation by providing a more energetic flux that enhances the capture probability, even under low-density conditions. Second, the range of energies affects the interactions between DM and stellar matter. At high energies, interactions may proceed through mechanisms such as deep inelastic scattering (DIS) or inelastic scatterings via the production of $N-$ and $\Delta-$ resonances.  At lower energies, interactions with nucleons or with nuclei become more relevant.  
The inclusion of BDM  complements conventional studies focused on low-energy regimes by incorporating DIS and resonant interactions. These varying interaction regimes will affect the capture process and must be accounted for when computing the capture rate probability. Furthermore, BDM enables the study of relativistic flux  contributions  and ensures that even non-relativistic interactions experience an effective enhancement, making it a powerful tool for exploring DM capture across all energy regimes.

 Our main goal is to explore the conditions under which  DM capture in WDs is most likely to occur. We do not focus on placing limits on DM interactions, since we would need to model the fluxes and do the study for specific cases, and this is beyond the scope of this work. 


To compute the DM interactions with stellar matter in such scenario, we move from a model-independent approach to focusing on DM fermions interacting through either a vector (dark photon) or dark scalar mediator. We will focus on the case of a single DM field, $\chi$, and examine interactions mediated by both the dark photon and the dark scalar separately.

To account for the interaction along different energy regimes, we have performed some computations for the first time in \Cref{sec:DM_cross-section} and particularly in \Cref{sec:Resonances}. We have used the relativistic theory of Fenyman, Kislinger and Ravndal (FKR) model~\cite{Feynman:1971wr} and an approach similar to Rein and Sehgal \cite{Rein:1980wg} to compute the inelastic interaction through resonant production of baryons for a massive current of DM fermions mediated by a vector and a scalar. The introduction of two massive particles (incoming and outgoing DM fermions) and a full propagator (dark photon or scalar) takes the generalization done by \cite{Berger:2007rq} to a next step. Therefore, the form factors and thereby the cross sections obtained in that section are completely original. Along with that, the treatment of the scalar interaction with nucleons in \Cref{sec:nucleons} is also based on an original approximation, in order to obtain simplified expressions for the scalar amplitudes.

The remainder of the paper is structured as follows. In \Cref{sec:DM_capture}, we briefly introduce the DM capture formalism and outline the main considerations when accounting for a wide range of energies. \Cref{sec:DM_cross-section} provides the  discussion of  DM interactions with stellar matter across different kinetic energy regimes. Our results are analysed in \Cref{sec:results}, and we conclude in \Cref{sec:conclusion}.

\section{Dark matter capture in compact stars}
\label{sec:DM_capture}

\subsection{Capture rate of DM}
When DM particles gravitate towards WDs, they interact with the constituents of stellar matter, leading to the potential capture of them within the star. Such interactions predominantly involve scattering off the nuclei and electrons of the stellar matter. With our study focusing specifically on the capture by DM scattering off nuclei elements, the capture rate of DM particles by a WD can be quantified using the expression~\cite{Busoni:2017mhe}:

\begin{equation}
\label{eq:capture_rate_gen}
C = \frac{\rho_\chi}{m_\chi}\int_0^{R\star}dr4\pi r^2\int_0^\infty du_\chi \frac{\omega}{u_\chi}f_{\rm MB}(u_\chi)\Omega^-(\omega),
\end{equation}
where $R_\star$ represents the radius of the WD, $\rho_\chi$ and $m_\chi$ denote the DM density and mass respectively, $u_\chi$ is the DM velocity at infinity, and $\omega$ is the DM velocity after it approaches the WD, at the point of interaction. Additionally, $f_{\rm MB}$ is the Maxwell-Boltzmann DM distribution function and $\Omega^-(\omega)$ represents the scattering probability of DM particles off stellar targets, a crucial aspect for our exploration of capture rates in WDs. As DM particles lose energy during these interactions, i.e., their velocity after scattering is lower than the escape velocity $v_e$, they gradually become trapped within the stellar structure, eventually concentrating at its core. To quantify this phenomenon, we express the DM interaction rate in terms of the differential cross-section function:
\begin{equation}
\label{eq:Omega_gen}
\Omega^-(\omega)=\int_0^{v_e} dv R^- (\omega\to v).
\end{equation}
Here $R^-$ is the differential interaction rate, denoting the rate at which a DM particle with velocity $\omega$ undergoes scattering with a target, resulting in a final velocity $v$. In the laboratory frame, this is determined by multiplying the differential cross-section, the Maxwell-Boltzmann density, and the relative velocity between a DM particle with velocity $\omega$ and a nucleus with velocity $v_T$~\cite{Gould:1987ir,Gould:1987ju},
\begin{equation}
    R^-(\omega\to v)= \frac{d\sigma}{dv}(\omega^2+v_T^2+2v_T\omega \cos\theta) n_T(r)\,2\pi \left(\frac{m_T}{2\pi T}\right)^{3/2} v_T^2 \, e^{-\frac{m_T}{2 T} v_T^2} dv_T\, d\cos\theta\label{Irate_general},
\end{equation}
where $\frac{d\sigma}{dv}$ is the differential cross section of the process, $\cos\theta$ is the scattering angle in the Lab frame.  $m_T$, $n_T$ and $T$ are respectively the mass, the  number density and the temperature of the target.  Observe that the Maxwell-Boltzmann distribution in the zero-temperature limit is a Dirac delta function,\footnote{
To formally verify this, we can express  the Maxwell-Boltzmann distribution in the following form,
$$
\lim_{T\to 0} \left(\frac{m_T}{2 T}\right)^{3/2} \left(v_T^2 \, e^{-\frac{m_T}{2 T} v_T^2}\right),
$$
in this expression, $\left(\frac{m_T}{2 T}\right)^{3/2}$ tends to infinity as $T$ approaches $0$. Additionally, the exponential also approaches zero and becomes increasingly narrow as $T\to0$.

This limit can be recognised as the definition of the Dirac delta function:
$$\delta(v_T) = \lim_{a \to \infty} \frac{1}{a} f\left(\frac{v_T}{a}\right),$$
where $f(v_T)$ is any well-behaved function. In this case, $f(v_T) = e^{-\frac{m_T}{2} v_T^2}$, and $a = \sqrt{T}$ (since $T$ approaches 0).
Therefore, in the limit of zero-temperature, the expression $\left(\frac{m_T}{2 T}\right)^{3/2} v_T^2 , e^{-\frac{m_T}{2 T} v_T^2}$ behaves like the Dirac delta function $\delta(v_T)$.
}
which simplifies the expression in Eq. \eqref{Irate_general}. After integrating over the target velocity using the delta-function and over the scattering angle, we obtain
\begin{equation}
      R^-(\omega\to v)= \frac{4}{\sqrt{\pi}} \frac{d\sigma}{dv} \omega^2 n_T(r),
\end{equation}
and the interaction rate given in the lab frame is 
\begin{equation}
    \Omega^-(\omega)=\frac{4}{\sqrt{\pi}} \int_0^{v_e} dv \frac{d\sigma}{dv} \omega^2 n_T(r).
\end{equation}
On the other hand, 
if we follow instead the computation in the centre of mass frame, the interaction rate in the limit of zero temperature results in \cite{Busoni:2017mhe},
\begin{eqnarray}
\Omega^-(\omega)&=& 
\frac{4\mu_+^2}{\mu \omega}n_T(r) \int_{\omega\frac{|\mu_-|}{\mu_+}}^{v_e}dv v \dfrac{d\sigma}{d\cos \theta},\label{eq:capture_prob_specific}
\end{eqnarray}
where  $\mu=m_\chi/m_T$, $\mu_\pm=(\mu\pm1)/2$.
%

\subsection{Capture rate density}
  
When considering a wide range of energies, including accelerated DM particles colliding with a WD, the capture rate described in \Cref{eq:capture_rate_gen} can change significantly. In the case of relativistic DM, the energy of the particles may reach levels at which the likelihood of capture becomes exceedingly low or even negligible. However, the much higher cross sections associated with these interactions can still contribute to visible capture rates, even for these high energies.

Since we are no longer interested in the conventional local DM flux, but rather in contributions from various sources, the standard Maxwell-Boltzmann distribution used to model local DM becomes invalid. Consequently, we should incorporate a DM flux in the expression for capture that accounts for these contributions. \response{We consider a velocity-flux with fixed energy, which is translated into a delta-distribution for a fixed velocity. This makes it possible to account for the probability to be captured at a particular energy, in order to have a map of those energies the star is sensitive to}.

In what follows, we leave the DM density as a free parameter and we define what we call the ``capture rate density"  as $\mathscr{C}=\rho_\chi^{-1}\, C$,
\begin{equation}
\label{eq:capture_rate_BDM}
\mathscr{C}=\frac{1}{m_\chi}\int_0^{R\star}dr4\pi r^2\int_0^\infty du^\prime_\chi \frac{\omega}{u^\prime_\chi}\delta(u^\prime_\chi - u_\chi)\,\Omega^-(\omega).
\end{equation}
This simplifies the expression and integrating over the DM velocity far away from the star $u_\chi$ is straightforward. Assuming the maximum capture probability, $\Omega^-(\omega) \to 1$, we obtain the geometric or optically thick limit
\begin{equation}
    \mathscr{C}_{\rm geom}=\frac{\pi R_\star^2}{m_\chi}\int_0^\infty du^\prime_\chi \frac{\omega}{u^\prime_\chi}\delta(u^\prime_\chi - u_\chi),
\end{equation}
where $\omega$ is the velocity of $\chi$ at the surface of the star. This quantity is independent of the DM interaction and allows us to determine up to what energies the capture of DM in WDs depends on the energy of the DM.

The second component we shall address in  our analysis is the computation of the interaction rate in \Cref{eq:Omega_gen}. This expression must encompass the implications of the model taken into account  and the associated high energies we want to explore. This means computing the scattering cross section within the model framework we are considering across different regimes, including DIS with quarks in the high-energy regime, possible resonances in the inelastic scattering process due to the high energies followed by elastic scattering with nucleons and nuclei. 

Finally, the relationship between the DM velocities near and far from the star is also modified due to the possible high velocities the BDM could possess. Considering a radial motion, this relationship becomes,\footnote{For more details on how we derived this relationship, see~\Cref{app:DM_velocities}}
\begin{equation}
    \label{eq:new_w_u_relation}
    \omega^2 = v_e^2+(1-v_e^2)u_\chi^2.
\end{equation}

 Such considerations are essential for comprehensively understanding the interactions between BDM and stellar targets of WDs, as well as exploring the probability of capture. 
 
The capture density framework established here serves as a first-order approximation to explore the parameter space of DM capture, particularly at high energies. By modelling the DM flux with a \response{fixed velocity}, this approach provides a simplified yet effective tool to evaluate whether capture is feasible under these conditions. While not a complete representation of the underlying physics, this approximation offers a valuable starting point for assessing DM interactions in extreme regimes, such as those involving DIS and resonant inelastic interactions.  Hence, in the following sections, we will dedicate a detail analysis on the computation of the scattering cross sections at the four possible scenarios.

\vspace{1cm}

\section{Dark matter scattering-cross section}
\label{sec:DM_cross-section}
We will separately consider a dark photon, $Z'$, or a scalar, $\Phi$, interacting with the SM and with DM fermions, $\chi$. We will just regard the vectorial couplings to the SM to be purely of vectorial nature, and not axial. The interaction terms have the following form:
\begin{equation}
\label{eq:Lagrangian}
\begin{split}
\mathcal{L}_{Z^\prime} &= g_{Z^\prime i} \overline{\psi}_\mathrm{SM}^i \gamma^\mu \psi_\mathrm{SM}^i Z^\prime_\mu + g_D \overline{\chi} \gamma^\mu (g_V^\chi - g_A^\chi \gamma^5) \chi Z^\prime_\mu,\\
\mathcal{L}_{\Phi} &= g_\Phi^{i} \overline{\psi}_\mathrm{SM}^i \psi_\mathrm{SM}^i \Phi + g_D \overline{\chi} \chi \Phi\,,
\end{split}
\end{equation}
where $\psi_\mathrm{SM}^i$ is a SM fermion. In our case, we are mainly interested in quarks, because the interactions will be computed with nuclei, nucleons and the quarks that belong to them.

With these considerations, in the following, we compute the DM scattering cross section in our two distinct scenarios: one involving interactions mediated by a dark photon, and the other by a dark scalar.

\subsection{Deep inelastic scattering}
\label{sec:DIS}
\begin{figure}
\centering
\begin{tikzpicture}[baseline=(current bounding box.center)]
\begin{feynman}
\vertex(V1);
\vertex(p1)[left=1.5cm of V1];
\vertex(p3)[right=1.5cm of V1];
\node[blob, below = 1.5cm of V1](V2);
\vertex(kXi)[below=1.3cm of p3];
\vertex(kXf)[below=2.5cm of p3];
\vertex(p2)[left=1.5cm of V2];
\diagram*{
(p1)--[fermion, edge label=\(\chi (p_1)\)](V1),
(p2)--[fermion, edge label'=\(N(p_2)\)](V2),
(V1)--[boson, edge label' = \(Z'/\Phi (q)\)](V2),
(V1) --[fermion, edge label=\(\chi (p_3)\)](p3),
(V2) --[fermion](kXi),
(V2) --[fermion](kXf),
};
\vertex(X1) [right=.2cm of kXi];
\vertex(X2) [right=.2cm of kXf];
\draw [decoration={brace}, decorate] (X1) -- (X2) node [pos=0.5, right = .2em] {\(X\)}  node [pos=0.4, left=.2em] {\(\vdots\)};
\end{feynman}
\end{tikzpicture}
\caption{Deep inelastic scattering mediated either by the dark photon or the dark scalar.}
\label{fig:DIS_feynman}
\end{figure}
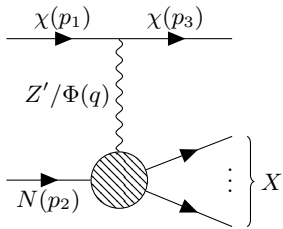

When the DM energies are high, well beyond the mass of the nucleons, the incoming fermions can interact directly with the quarks that constitute the nucleons: the valence quarks, which are $uud$ for the proton and $udd$ for the neutron, as well as the sea quarks that become visible at high energies. Due to the property of asymptotic freedom, these quarks can interact freely, behaving as if their binding with other quarks were negligible. They carry a fraction $\xi$ of the total momentum of the nucleon, and under this approach, they are also known as \textit{partons}. This interaction leads to the production of a hadronic shower, and the scattering is far from elastic; instead  it is a deep inelastic scattering (DIS). In \Cref{fig:DIS_feynman}, we show the diagram corresponding to this process where the interaction is mediated either by a dark photon $Z^\prime$ or a scalar $\Phi$. 
%
%
Although at these high energies it becomes more difficult to produce outgoing fermions below the escape velocity of WDs, reducing the likelihood of capture, the higher cross sections increase the interaction rate. This makes the energy region particularly interesting, as the capture process becomes highly efficient. Additionally, the high production of particles as hadronic jets in DIS opens a rich phase space where a damped outgoing DM particle remains possible.

At these very high energies, there are some quantities that are important for computing the cross section of this process,
%
\begin{itemize}
    \item \textit{Momentum transfer}: $Q^2 \equiv -q^2$
    
    \item \textit{Energy transfer}: $\nu \equiv \frac{p_2\cdot q}{m_N}$, such that $m_N$ is the mass of the nucleon.
    
    \item \textit{Inelasticity}: $y \equiv \frac{p_2\cdot q}{p_2\cdot p_1}$
    
    \item \textit{Bjorken scaling variable}: $x \equiv \frac{Q^2}{2 p_2\cdot q}$.
\end{itemize}
%
This holds for both, vector and scalar mediator scenarios. The Bjorken scaling  variable $x$ coincides at high energies with the momentum fraction $\xi$ taken by each quark or parton in the computation. The cross section will be found by integrating on the $x-y$ plane, instead of $E_3$ and $\cos \theta$, where $E_3$ is the energy of the scattered DM and $\theta$ is its  angle with respect to the incoming one in the Lab frame. The relation between those is straightforward,
\begin{align}
E_3 &= (1-y) E_\chi,\\
\cos\theta &= 1 - \frac{m_N x y}{E_\chi (1-y) },
\end{align}
such that $E_\chi$ is the energy of the incoming fermion. 
%

\subsubsection{DIS with a dark photon}
Accounting for a dark photon as the mediator of the interaction $\chi q \to \chi X$, where the quark $q$ carries a fraction $\xi = x$ of the nucleon momentum, results in the following squared partonic amplitude,
\begin{equation}
\begin{split}
\frac{1}{4} \sum_s |\mathcal{M}|^2 = &8 g_D^2 g_{Z^\prime q}^2\frac{E_\chi^2 m_N^2 x^2}{m_{Z^\prime}^4 (1 + Q^2 / m_{Z^\prime}^2 )^2} \\&\times \left( \big( A_q A_\chi + C_q C_\chi \big) y^2 - 2 \big( A_q A_\chi - C_q C_\chi \big) y - \frac{A_q m_\chi^2}{E_\chi m_N x} B_\chi y + 2 A_q A_\chi \right).
\end{split}
\end{equation}
Here, $s$ represents the initial spins, $m_{Z^\prime}$ denotes the dark photon mass, and $A_k \equiv \big(g^k_V\big)^2 + \big(g^k_A\big)^2 $, $B_k \equiv \big(g^k_V\big)^2 - \big(g^k_A\big)^2 $, and $C_k \equiv 2 g_k^V g_k^A $ are computed from the vector and axial coefficients of the quark ($k = q$) and DM ($k = \chi$) currents. For the present case, the axial coefficients of the quarks vanish. We have considered four initial spin configurations.

To compute the differential cross section, we use the Fermi Golden Rule, incorporating the factor arising from the partonic distribution \( f_q(\xi) \), which gives the probability of finding a parton (quark) carrying the fraction $\chi$ of the nucleon’s longitudinal momentum during the DIS process.

After integrating all parameters while retaining the masses of the DM particles, we derive the DIS cross section
\begin{equation}
\begin{split}
\frac{d^2\sigma}{dx dy} = & \frac{g_D^2}{4 \pi m_{Z^\prime}^4} \frac{E_\chi^2 m_N x}{ (1 + Q^2 / m_{Z^\prime}^2 )^2} \frac{\sqrt{E_\chi^2 (1 - y)^2 - m_\chi^2}}{(1-y)(E_\chi^2 - m_\chi^2)} \\&\times \sum_q g_{Z^\prime q}^2 \left( \big( A_q A_\chi + C_q C_\chi \big) y^2 - 2 \big( A_q A_\chi - C_q C_\chi \big) y - \frac{A_q m_\chi^2}{E_\chi m_N x} B_\chi y + 2 A_q A_\chi \right) f_q(x).
\end{split}
\end{equation}
Notice that setting $m_\chi$ to zero yields the conventional formula for DIS neutral current interactions with fermions.

\subsubsection{DIS with a dark scalar}
The interaction $\chi q \to \chi X$ with a scalar mediator $\Phi$ yields the following squared amplitude,
\begin{equation}
\begin{split}
\frac{1}{4} \sum_s |\mathcal{M}|^2 = &g_D^2 g_{\Phi q}^2 \frac{\big(4 m_\chi^2 + Q^2\big)\big(2 m_N^2 x^2 + Q^2\big)}{(m_\Phi^2 + Q^2)^2}.
\end{split}
\end{equation}
Where once more, we have set $\xi=x$ and we do not neglect the masses.

For the computation of the DIS differential cross section with a scalar mediator, we need to incorporate the squared amplitude into the calculation by integrating over the parton distribution within the nucleons.
The final expression for the differential cross section is given by,
\begin{equation}
\begin{split}
\frac{d^2\sigma}{dx dy} = & \frac{g_D^2 }{16 \pi m_\Phi^4} \frac{y E_\chi^2 (Q^2 + 4 m_\chi^2)}{(1 + Q^2/m_\Phi^2)^2 (E_\chi^2 - m_\chi^2)} \times \sum_q g_{\Phi q}^2 f_q(x).
\end{split}
\end{equation}

 
\subsection{Resonant scattering}
 \label{sec:Resonances}
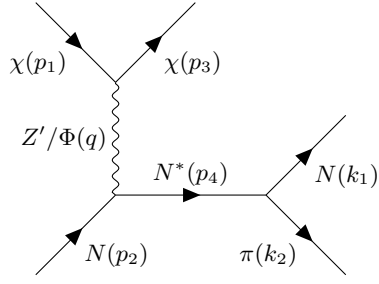
\begin{figure}
\centering
\begin{tikzpicture}[baseline=(current bounding box.center)]
\begin{feynman} 
\vertex(V1);
\vertex(p1)[above left=1.5cm of V1];
\vertex(p3)[above right=1.5cm of V1];
\vertex[below = 1.5cm of V1](V2);
\vertex(p4)[right=2cm of V2];
\vertex(p2)[below left=1.5cm of V2];
\vertex[right=2cm of V2](V3);
\vertex(p5)[above right=1.5cm of V3];
\vertex(p6)[below right=1.5cm of V3];
\diagram*{
(p1)--[fermion, edge label'=\(\chi(p_1)\)](V1),
(p2)--[fermion, edge label'=\(N(p_2)\)](V2),
(V1)--[boson, edge label'=\(Z^\prime / \Phi(q)\)](V2),
(V1) --[fermion, edge label'=\(\chi(p_3)\)](p3),
(V2) --[fermion, edge label=\(N^*(p_4)\)](p4),
(V3) --[fermion, edge label'=\(N(k_1)\)](p5),
(V3) --[fermion, edge label'=\(\pi(k_2)\)](p6)
}; 
\end{feynman}
\end{tikzpicture}
\caption{DM-nucleon inelastic interaction with a dark photon ($Z'$)/scalar ($\Phi$) mediator that produces a $N-$ or $\Delta-$ resonance ($N^*$) that further decays into a nucleon and a pion.}\label{diagram:DM_RES}
\end{figure}

The resonant scattering refers to an inelastic interaction with a nucleon that produces a $N-$ and $\Delta-$resonance nearly on-shell that further decays into a nucleon and a pion, as can be seen in \Cref{diagram:DM_RES},
\begin{equation}
    \chi + N \to \chi + N^* \to \chi + N + \pi.
\end{equation}
In the case of neutral mediators like the dark photon or scalar, there are four possible channels:
\begin{enumerate}
    \item $\chi + p \to \chi + p + \pi^0$,
    \item $\chi + p \to \chi + n + \pi^+$,
    \item $\chi + n \to \chi + n + \pi^0$,
    \item $\chi + n \to \chi + p + \pi^-$.
\end{enumerate}

The approach we take here is similar to what Rein and Sehgal did for neutrinos in Ref.~\cite{Rein:1980wg}. The authors considered a similar interaction as ours, but instead of the dark matter current they regarded a massless leptonic current composed of either two neutrinos or a neutrino with a charged lepton and a Fermi effective interaction. A massive charged lepton was considered by~\cite{Kuzmin:2003ji}, and this work was later used in~\cite{Berger:2007rq} to adapt Rein and Sehgal's original cross section computation to include one massive lepton and the effects of a pion pole term in the axial amplitude. This change gave rise to two different contributions depending on the helicity of the charged massive lepton. We have taken a third step, so that the DM current contains two massive particles that give rise to four helicity combinations and a mediator whose propagator is not independent of the momentum transfer in general.\footnote{For an extended version of the calculations presented in this section, please refer to \Cref{sec:App_resonances}. } For more information about the modelling of the hadron currents and amplitudes, we refer the reader to the papers cited just above and \cite{Adler:1968tw, Feynman:1971wr, Ravndal:1973xx}. For a full treatment of the cross sections and amplitudes in BSM scenarios, we refer the reader to \cite{HoefkenZink:2025tns}.

%
%
\subsubsection{Dark photon mediator}

The amplitude in the case of a dark photon mediator is as follows and depends on the helicities of the DM particles:

\begin{equation}
\begin{split}
&\mathcal{M} (\chi(p_1, \lambda_1) N(p_2) \to \chi(p_3, \lambda_2) N^{*} (p_4))\\ &= \frac{g_D g_{Z^\prime N}}{q^2 - m^2_{Z^\prime}} \left[ \overline{u}_{p_3 \lambda_2} \gamma_\mu \left( g_V - g_A \gamma^5 \right) u_{p_1 \lambda_1} \right] \left( g^{\mu \nu} - q^\mu q^\nu / m^2_{Z^\prime}  \right) \bra{N^{*}} J_\nu^{+} (0) \ket{N} \\
&= 2M \frac{g_D g_{Z^\prime N}}{q^2 - m^2_{Z^\prime}} V_{\lambda_1 \lambda_2}^\mu \bra{N^{*}} F_\mu^{V} \ket{N},
\end{split}
\end{equation}
where $g_{Z^\prime N}$ represents the order of the couplings with the SM, such that $g_{Z^\prime N} g^V_q$ would be the full vectorial coupling to quark $q$, and $V_{\lambda_1 \lambda_2}^\mu \equiv \left[ \overline{u}_{p_3 \lambda_2} \gamma_\nu \left( g_V - g_A \gamma^5 \right) u_{p_1 \lambda_1} \right] \left( g^{\mu \nu} - q^\mu q^\nu / m^2_{Z^\prime}  \right)$ represents the vector interacting with the hadronic current. This vector can be decomposed into components associated with different helicity states:

\begin{equation}
V_{\lambda_1 \lambda_2}^\mu = C_L^{\lambda_1 \lambda_2} e_L^\mu + C_R^{\lambda_1 \lambda_2} e_R^\mu + C_{s}^{\lambda_1 \lambda_2} e_{\lambda_1 \lambda_2}^\mu,
\end{equation}
where $e_L^\mu$, $e_R^\mu$, and $e_{\lambda_1 \lambda_2}^\mu$ are basis vectors depending on the helicities. The coefficients $C_k^{\lambda_1 \lambda_2}$, for $k = R, L, s$, which stand for right, left and scalar (longitudinal) polarizations respectively, are defined by:

\begin{equation}
C_k^{\lambda_1 \lambda_2} = \frac{e^*_{k,\mu} V^{\mu}_{\lambda_1 \lambda_2}}{e^*_{k,\nu} e_k^\nu}.
\end{equation}
The vector current $J_\nu^{+} (0)$ is expressed as $2 M F_\mu^{V}$, with $M$ being the mass of $N^*$. 

The differential cross section can be computed straightforwardly in terms of two parameters: the transferred momentum $q^2$ and the invariant mass $W \equiv \sqrt{p_4^2}$. We will compute each kind of transition $k = R, L, s$ separately and we also need to sum all possible combinations for $\lambda_1 \lambda_2$. The result is expressed in terms of the helicity cross sections $\sigma_k$,

\begin{equation}
\begin{split}
\frac{d\sigma}{d W d q^2} &= \frac{\alpha g_D^2 g_{Z^\prime N}^2}{\pi \left( q^2 - m_{Z^\prime}^2 \right)^2} \frac{W}{m_N} \sum_{\lambda_1 \lambda_2} \left[  \left|C_L^{\lambda_1 \lambda_2} \right|^2 \sigma_L + \left|C_R^{\lambda_1 \lambda_2} \right|^2 \sigma_R + \left|C_{s}^{\lambda_1 \lambda_2} \right|^2 \sigma_s^{\lambda_1 \lambda_2} \right],
\end{split}
\end{equation}
where $\alpha \equiv e^2 / (4\pi)$ is the fine structure constant.

The delta distribution function $\delta \left( W - M \right)$, where $M$ is the resonance mass, accounts for the narrow width in which the resonance takes place. We assume that this width is finite, so we need to replace the delta by a Breit - Wigner factor,

\begin{equation}
\delta \left( W - M \right) \to \frac{1}{2\pi} \frac{\Gamma}{(W - M)^2 + \Gamma^2 / 4}.
\end{equation}

The details on how to compute $\Gamma$ can be found in \cite{Rein:1980wg}, together with the tables of data. We have also used the more updated data from \cite{Kabirnezhad:2017xzx}. The list of resonances can also be found in \cite{Rein:1980wg, Kabirnezhad:2017xzx}. The values of each of their amplitudes can be computed from Table I in \cite{HoefkenZink:2025tns}. The parameters used to compute these amplitudes are similar to those of Rein and Sehgal~\cite{Rein:1980wg}, we just need to extend the definition of the parameter $S$ to account for the different combinations of helicities,
\begin{equation}
\label{eq:S_definition}
S \to S^{\lambda_1 \lambda_2} \equiv \frac{\sqrt{-q^2}}{\left|\vec{q}_\mathrm{Lab}\right|^2} \frac{V_{\lambda_1 \lambda_2}^3 q^0_\mathrm{IB} - V_{\lambda_1 \lambda_2}^0 \left|\vec{q}_\mathrm{IB}\right|}{C_s^{\lambda_1 \lambda_2}} \left(\frac{1}{6} - \frac{q^2}{6 m_N^2} - \frac{W}{2m_N} \right) F^V (q^2)
\end{equation}
where $\mathrm{IB}$ refers to the Isobaric frame, well described in \cite{Kabirnezhad:2017xzx} and the vector dipole function $F^V$ is taken from \cite{Graczyk:2007bc}. For further details on the computation of the resonant scattering cross section, please refer to Appendix~\ref{sec:App_resonances}.

Finally, it is important to consider interferences in the computation of the different resonances. Those that share the same orbital ($L$) and total ($J$) angular momentum interfere: so before squaring the partial results it is important to sum resonances of the form $L_{2I,2J}$, following the FKR nomenclature convention~\cite{Feynman:1971wr}, where $L$ and $J$ are the same, but the isospin $I$ not necessarily. Each isospin state must be accompanied by its Clebsch - Gordan factor according to FKR~\cite{Feynman:1971wr}, on which the Rein - Sehgal approach is based. We refer the reader to those papers, since going into details is beyond the scope of the present paper.

\subsubsection{Scalar mediator}

For the scalar mediator case, we have extended the Rein-Sehgal model to accommodate this new mediator. This work presents, for the first time, the computation of the scalar-mediated resonance cross section using this approach, which is explained in the following lines. 
In order to facilitate notation, we will consider the couplings to each quark to be: $g_{\Phi q} \equiv g_{\Phi N} \times g^S_q$, where $g^S_q$ would play the role of a vectorial or axial coefficient but for scalars, while $g_{\Phi N}$ gives the overall order of the coupling. The amplitude of the process is,
\begin{equation}
\label{eq:res_scalar_amplitude}
\begin{split}
\mathcal{M} (\chi(p_1, \lambda_1) N(p_2) \to \chi(p_3, \lambda_2) N^{*} (p_4)) &= \frac{g_D g_{\Phi N}}{q^2 - m^2_{\Phi}} \left[ \overline{u}_{p_3 \lambda_2}u_{p_1 \lambda_1} \right]  \bra{N^{*}} J_S^{+} (0) \ket{N} \\
&= 2M \frac{g_D g_{\Phi N}}{q^2 - m^2_{\Phi}} V_{\lambda_1 \lambda_2}^S \bra{N^{*}} F_S \ket{N},
\end{split}
\end{equation}
where $J_S^{+}$ is the scalar current associated with the baryons and $F_S \equiv 2 M J_S^{+} (0)$ the form factor associated to it, and $V_{\lambda_1 \lambda_2}^S \equiv \left[ \overline{u}_{p_3 \lambda_2}u_{p_1 \lambda_1} \right]$. We can compute the cross section in the same way we did for the vector mediator:

\begin{equation}
\begin{split}
\frac{d\sigma}{d W d q^2} &= \frac{g_D^2 g_{\Phi N}^2}{64 \pi \left( q^2 - m_\Phi^2 \right)^2} \frac{W}{m_N^2 |\vec{p}_1|^2} \sum_{\lambda_1 \lambda_2} |V_{\lambda_1 \lambda_2}^S|^2 \sum_{i = \pm} \bra{N^{*}} F^S_{0^\pm} \ket{N}.
\end{split}
\end{equation}
Here $\bra{N^{*}} F^S_{0^\pm} \ket{N}$ is the form factor for the transition from $\ket{N, \pm 1/2}$ to $\ket{N^{*}, \pm 1/2}$, where we are showing the spin explicitly. It is independent of the helicities on the DM current. You can see the appendix for more details about the computation of the form factor within the FKR model. The kinematics of this interaction is exactly the same as in the dark photon mediator case. It is important to note that the resonant scattering through a scalar is merely an estimation based on the FKR model, which would need to be validated through experiments. However, this is not feasible due to the experimentally unexplored nature of the particles involved: DM. In the axial case, for instance, the authors introduced a normalisation factor of $0.75$ to accommodate the form factor to its experimental value at $q^2 = 0$~\cite{Rein:1980wg}.


\subsection{Elastic scattering on nucleons}
\label{sec:nucleons}
\begin{figure}
\centering
\begin{tikzpicture}[baseline=(current bounding box.center)]
\begin{feynman} 
\vertex(V1);
\vertex(p1)[above left=1.5cm of V1];
\vertex(p3)[above right=1.5cm of V1];
\vertex[below = 1.5cm of V1](V2);
\vertex(p4)[below right=1.5cm of V2];
\vertex(p2)[below left=1.5cm of V2];
\diagram*{
(p1)--[fermion, edge label'=\(\chi(p_1)\)](V1),
(p2)--[fermion, edge label=\(N(p_2)\)](V2),
(V1)--[boson, edge label'=\(Z^\prime / \Phi (q)\)](V2),
(V1) --[fermion, edge label'=\(\chi(p_3)\)](p3),
(V2) --[fermion, edge label=\(N(p_4)\)](p4)
}; 
\end{feynman}
\end{tikzpicture}
\caption{DM-nucleon interaction with a dark photon ($Z'$)/scalar ($\Phi$) mediator.}\label{diagram:DM_nucleons}
\end{figure}
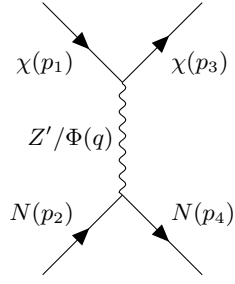

When the energies of the incoming $\chi$ are lower than those required for DIS, the possibility of direct interaction with the quarks diminishes, leading to a focus on elastic interactions with nucleons instead. To understand these interactions, it becomes essential to introduce form factors. In the following, we will briefly discuss this formalism to facilitate the computation of such interactions, as depicted by the diagram in \Cref{diagram:DM_nucleons}, where the interaction is mediated either by a dark photon or a dark scalar. The approach is well known in the literature~\cite{walecka2001electron, gao2003nucleon, hyde2004electromagnetic, thomas2010structure}.
%
%
\subsubsection{DM - nucleon interaction with a dark photon}

The general expression for the amplitude of the interaction depicted in Fig.~\ref{diagram:DM_nucleons} mediated by a dark photon is given by,
\begin{equation}
\begin{split}
\mathcal{M}_N &= i \frac{g_D g_{Z^\prime N}}{q^2 - m_{Z^\prime}^2} [\overline{u}(p_3) \gamma^\mu (g_V^\chi - g_A^\chi \gamma^5) u(p_1)] \left(g_{\mu \nu} - \frac{q_\mu q_\nu}{ m_{Z^\prime}^2}\right) \bra{N(p_4)} j_{Z^\prime Q}^\nu (0)\ket{N(p_2)}
\end{split}
\end{equation}
where $N$ represents the nucleon ($p$ or $n$), and $g_\mathbf{Had}$ denotes a general order of the couplings of the dark photon with the quarks. This coupling accounts for the differences in how the dark photon couples with different quarks, which are absorbed by the vector and axial coefficients in the hadronic current, defined as:
\begin{equation}
\begin{split}
j_{Z^\prime Q}^\nu  &= \sum_q g_V^q \overline{q}\gamma^\nu q - \sum_q g_A^q \overline{q}\gamma^\nu \gamma^5 q.
\end{split}
\end{equation}

There is a dependence on the position, $x$, implicit in the previous equation. As observed, this hadronic current can be decomposed into a vectorial and an axial part: $j_{Z^\prime Q}^\nu \equiv v_{Z^\prime Q}^\nu - a_{Z^\prime Q}^\nu$, such that,
\begin{align}
\label{eq:vector_axial_coefficients}
v_{Z^\prime Q}^\nu  &= -2(g_V^u + 2 g_V^d) v_3^\nu + 3(g_V^u + g_V^d) j_{A Q}^\nu + (g_V^u + g_V^d + g_V^s) v_s^\nu \nonumber\\ &\quad - [g_V^s \overline{b} \gamma^\nu b + (3g_V^u + 3g_V^d + g_V^s) (\overline{c} \gamma^\nu c + \overline{t} \gamma^\nu t)]\\
a_{Z^\prime Q}^\nu  &= (g_A^u - g_A^d) a_3^\nu + (g_A^u + g_A^d) a_0^\nu + g_A^s a_s^\nu - \sum_{q=c,b,t} (g_A^s - g_A^q) \overline{q} \gamma^\nu \gamma^5 q.
\end{align}
 For further details, please refer to Appendix \ref{Ap:FF_details}. The hadronic currents would then be,
\begin{align}
\bra{N(p_4)} v_{Z^\prime Q}^\mu (0)\ket{N(p_2)} &= \overline{u}_N (p_4) \bigg[ \gamma^\mu F_1^{Z^\prime N} (Q^2) + i\frac{q_\nu}{2m_N} \sigma^{\mu \nu} F_2^{Z^\prime N} (Q^2) \bigg] u_N (p_2), \\
\bra{N(p_4)} a_{Z^\prime Q}^\mu (0)\ket{N(p_2)} &= \overline{u}_N (p_4) \bigg[ \gamma^\mu \gamma^5 G_A^{Z^\prime N} (Q^2) + \frac{q_\mu}{m_N} \gamma^5 G_P^{Z^\prime N} (Q^2) \bigg] u_N (p_2).
\end{align}
If the contribution from currents associated with the quarks $c$, $b$, and $t$ is negligible, and if we define the axial form factors $G_k^{0 N}$ as those arising from the contribution of the axial current $a_0^\mu$, then the overall form factors would be as follows,
\begin{align}
F_i^{Z^\prime N} &\simeq \mp (g_V^u + 2g_V^d)(F_i^p - F_i^n) + 3(g_V^u + g_V^d) F_i^N + (g_V^u + g_V^d + g_V^s) F_i^{sN} \\
G_k^{Z^\prime N} &\simeq \pm \frac{1}{2} (g_A^u - g_A^d) G_k + (g_A^u + g_A^d) G_k^{0 N} + g_A^s G_k^{sN},
\end{align}
where $N = p, n$, $i = 1,2$ and $k = A, P$. The remaining form factors are the well-known ones that appear in the SM.
A particularly simple case is when the vectorial interaction is proportional to that of the photon, the vectorial coefficients are proportional to the electromagnetic charges and the axial ones vanish, leading to $F_i^{Z^\prime N} = F_i^N$ and $G_k^{Z^\prime N} = 0$. Consequently, the amplitude would be
\begin{equation}
\begin{split}
\mathcal{M}_N &= i \frac{g_D g_{Z^\prime N}}{q^2 - m_{Z^\prime}^2} [\overline{u}(p_3) \gamma^\mu (g_V^\chi - g_A^\chi \gamma^5) u(p_1)] \left(g_{\mu \nu} - \frac{q_\mu q_\nu} {m_{Z^\prime}^2}\right)\\ &\quad \times \overline{u}_N (p_4) \bigg[ \gamma^\nu F_1^{N} (Q^2) + i\frac{q_\lambda}{2m_N} \sigma^{\nu \lambda} F_2^{N} (Q^2) \bigg] u_N (p_2).
\end{split}
\end{equation}
To compute the full form of the DM-nucleon scattering amplitude, we employ the Sachs electric and magnetic form factors which are expressed as~\cite{Sachs:1962pwa}
\begin{align}
G_E^N (Q^2) &= F_1^N (Q^2) - \frac{Q^2}{4 m_N^2} F_2^N (Q^2) = \delta_{N p} G_D (Q^2), \\
G_M^N (Q^2) &= F_1^N (Q^2) + F_2^N (Q^2) = \frac{\mu_N}{\mu_\mathcal{N}} G_D (Q^2), \\
G_D (Q^2) &= (1 + Q^2 / m_V^2)^{-2}.
\end{align}
Here, $\mu_\mathcal{N} \equiv \frac{e \hbar}{2 m_p}$ represents the nuclear magneton, $\mu_N$ denotes the magnetic moments of the nucleons (proton and neutron), which are equal to $2.79 \mu_\mathcal{N}$ for the proton and $-1.91 \mu_\mathcal{N}$ for the neutron. $m_V$ is an experimental value that fits the dipole function $G_D$ and is approximately equal to $0.84$ GeV. With these quantities, we can compute the full cross section, which vanishes for the neutron because it does not interact through a vector whose interaction is proportional to the electromagnetic one. Additionally, when computing the cross section, we must consider the $4$ possible initial spin states and sum over all of them to obtain the mean value. Given that the expression for the amplitude is not simple, we do not present its analytical expression here.

\subsubsection{DM - nucleon interaction with a scalar}
To derive the interaction between DM and nucleons mediated by a scalar, we consider the effective vector current for nucleons, focusing on the contributions from $u$ and $d$ quarks. For energies below DIS and assuming an elastic collision, the transition element is computed using the nucleon currents as described by the effective vector current, detailed in \Cref{eq:nucleon_currents}. The scalar current is defined as: 
\begin{equation} 
j_S^\mu = \frac{g_{\Phi u}}{2 m_u} \overline{u} \gamma^\mu u + \frac{g_{\Phi d}}{2 m_d} \overline{d} \gamma^\mu d + r^\mu 
\end{equation} 
where $r^\mu$ is a residual current that depends on quarks other than the up or the down one. After contracting the expression with the sum of the initial and final momenta, we obtain the following transition amplitude, whose more detailed obtention can be found in \Cref{sec:app_dm_nucleon}:
\begin{equation} 
\begin{split} 
(p_{i \mu} + p_{f \mu}) \bra{N(p_f)} j_{S}^\mu (0) \ket{N(p_i)} &\simeq \overline{u}N (p_f) \left[2 m_N F_1^{S N} (Q^2) - \frac{Q^2}{2 m_N} F_2^{S N} (Q^2) \right] u_N(p_i)\\
&\equiv \bra{N(p_f)} g_{\Phi u} \overline{u} u + g_{\Phi d} \overline{d} d \ket{N(p_i)}. 
\end{split} 
\end{equation}
The second line on the RHS follows from a rough estimation that helps us in having an idea of how the scalar form factors could behave. When each of the momenta is applied to the current, we could consider the quarks are carrying a fraction $\xi_i$ for each quark $i$ of the total momentum, as in the parton model. Although that would need high energies, we will assume it to have an idea of the result. So the action of the momenta inside the integral of the current would contain the following factor:
\begin{equation}
\begin{split}
(p_{i \mu} + p_{f \mu}) \frac{g_{\Phi q}}{2 m_q} \overline{q} (\xi_1 p_{i}) \gamma^\mu q (\xi_2 p_{f}) &= \frac{g_{\Phi q}}{2 m_q} \overline{q} (\xi_1 p_{i}) (\frac{1}{\xi_1} \xi_1 \slashed{p}_{i} + \frac{1}{\xi_2} \xi_2 \slashed{p}_{f}) q(\xi_2 p_{f})\\
&= \frac{g_{\Phi q}}{2 m_q} \overline{q} (\xi_1 p_{i}) (\frac{1}{\xi_1} m_q + \frac{1}{\xi_2} m_q q(\xi_2 p_{f})\\
&\geq g_{\Phi q} \overline{q} q\,.
\end{split}
\end{equation}
We are assuming the main region of integration of $j_S^\mu (0)$ comes from $\xi_i \le 1$, as for the high energy regimes. Therefore, the action of the momenta, $(p_{i \mu} + p_{f \mu})$, on $\overline{q} \gamma^\mu q$ as $ 2 m_q \overline{q} q$ is a conservative approach: the minimum one would expect under the consideration of a partonic modelling. Since this is just a rough estimation, we will stick to it, because a more careful derivation is out of the scope of the present work.\footnote{For low energies with respect to the mass of the nucleons, it can also be shown that $ 2 m_q \overline{q} q$ is a lower bound for  that  $(p_{i \mu} + p_{f \mu}) \equiv p_\mathrm{i+f,\mu}$, on $\overline{q} \gamma^\mu q$ and that it can be used as a conservative choice. In the integral that contains the factor $\overline{q} (p_1) \slashed{p}_\mathrm{i+f} q (p_2)$, we can perform the integral such that $p_1$ always points to the $z$ axis and $p_2$ lies on the $XZ$ plane. Any term linear in the space indices of $p_\mathrm{i+f}$ should be negligible with respect to the other terms, that contain the total energies or the masses. Whether considering positive or negative helicities, the previous factor should be: $\left( \sqrt{(E_1 + m_q)(E_2 + m_q)} + \sqrt{(E_1 - m_q)(E_2 - m_q)} \right) E_\mathrm{i+f} \cos \frac{\theta}{2}$, where $\theta$ is the angle between the two quarks. The computation that replaces $\slashed{p}_\mathrm{i+f}$ by $2 m_q$ results in: $\left( \sqrt{(E_1 + m_q)(E_2 + m_q)} - \sqrt{(E_1 - m_q)(E_2 - m_q)} \right) 2 m_q \cos \frac{\theta}{2}$. Since the square roots are always positive and the energies of the incoming and outgoing nucleons are greater than twice the mass of any quark ($u$ or $d$), the latter expression is less than the former one. Therefore, it can be conservatively replaced.}

Using the scalar form factors $F_i^{SN}$, the differential cross section in the center-of-mass (CM) frame is given by: %
\begin{equation}
\label{eq:DM_N_Sxsec}
\frac{d\sigma^N}{dz} = \frac{g_D^2 g_{\Phi N}^2}{8 \pi m_N^2 \left(E_1+E_2\right){}^2} \frac{\left(E_1^2 + m_{\chi }^2 - p_1^2 z\right) \left(p_1^2 (1+z) +2 m_N^2\right) \left(2 F_1^{\text{SN}} m_N^2-p_1^2 F_2^{\text{SN}} (1-z)\right)^2}{\left(2 p_1^2 (1-z) + m_{\Phi}^2\right)^2} 
\end{equation}
where $z = \cos \theta$, $p_1 = \sqrt{E_1^2 - m_\chi^2}$ is the momentum of the incoming DM particle, $p_2=\sqrt{E_2^2-m_p^2}$ is the momentum of the incoming nucleon and $g_{\Phi N}$ and $g_{\Phi q}$ are the couplings to the nucleons and quarks, respectively. The full cross section is obtained by integrating this differential cross section.

For further details on the derivation and computation of the scalar current, transition amplitude, and form factors, please refer to~\Cref{sec:app_dm_nucleon}. Additionally, 
in \Cref{Ap:Helm_FF_computation}, we offer a concise discussion of computing the differential cross-section using an alternative approach, wherein we consider the well-established Helm form factors.


\subsection{Elastic Scattering on Nuclei}
\label{sec:nuclei}

The interaction can also take place with the whole nucleus. The diagram would be the same as in \Cref{diagram:DM_nucleons}, but with $N$ representing the nucleus instead of a nucleon. There are several studies attempting to model general interactions between DM and specific nuclei~\cite{Fitzpatrick:2012ix, Cirelli:2013ufw, Catena:2015uha}. In particular, \cite{Catena:2015uha} offers a comprehensive treatment of 15 types of non-relativistic operators, which describe the interaction between any DM candidates and any type of nucleus. Here, we primarily rely on these references to translate the interactions of \Cref{eq:Lagrangian} with nuclei into these structures in order to compute the full cross section.

The nuclear operators are described in terms of quantum operators constructed from spins and spatial kinematic quantities. Specifically, we consider $\vec{v} \equiv \vec{p}_\chi / m_\chi - \vec{p}_N / m_N$, representing the initial relative velocity between the DM particle and the nucleon, and $\vec{q}$, which denotes the transverse 3-momentum from the nucleon to the DM particles. From $\vec{v}$, we can derive the component of $\vec{v}$ perpendicular to $\vec{q}$, denoted as $\vec{v}^\perp$. With these quantities, we construct five Hermitian operators: $\mathrm{Id}_{\chi N}$, $i \hat{q} / m_N$, $\hat{v}^\perp$, $\hat{s}_\chi$, and $\hat{s}_N$, all of which are Galilean invariant.

Considering the 15 nuclear operators found in \cite{Catena:2015uha}, our objective is to compute the scattering amplitude given by,
\begin{equation}
\bra{\Psi_f} H_T \ket{\Psi_i} = (2 \pi)^3 \delta^{(3)} (\vec{p}_1 + \vec{p}_2 - \vec{p}_3 - \vec{p}_4) i \mathcal{M}_T^{NR}.\label{eq:scattering_amp}
\end{equation}
Here, $H_T$ is the full Hamiltonian, obtained by integrating the Hamiltonian density,
\begin{equation}
\mathcal{H}_T (\vec{r}) = \sum_{i=1}^{A} \sum_{j = 0,1} \sum_{k=1}^{15} c_k^j \mathcal{O}_k^{(i)} (\vec{r}) t^j{(i)},
\end{equation}
where $t^0{(i)} = \mathrm{Id}_2$ and $t^1{(i)} = \tau_3$ denote the third Pauli matrix for the $i$th nucleon. The constants $c_k^0$ and $c_k^1$ represent the isoscalar and isovector coupling constants, respectively, given by $c_k^0 \equiv c_k^p + c_k^n$ and $c_k^1 \equiv c_k^p - c_k^n$. These constants are the proton and neutron couplings scaling the operators $\mathcal{O}_k$ under the assumption of isospin invariance. The sum over $i$ encompasses all nucleons in the target nucleus.

Following \cite{Catena:2015uha}, the mean squared of the scattering amplitude equals,
\begin{equation}
\frac{1}{N_i}\sum_{i,j} \big| \mathcal{M}_{T}^{NR} \big|^2 = \frac{m_T^2}{m_N^2} \sum_{i, j}^{15} \sum_{\alpha, \beta = 0, 1} c_i^\alpha c_j^\beta F_{ij}^{\alpha \beta} (v^2, q^2, y).
\end{equation}
Here, $N_i$ represents the number of initial states, while $\alpha$ and $\beta$ iterate over the different isoscalar and isovector states. $F_{ij}^{\alpha \beta} (v^2, q^2, y)$ denotes the form factors accounting for the DM and nuclear response functions, and $y$ is a dimensionless variable given by $b^2 q^2 / 4$, where $b$ is the length parameter.\footnote{This parameter is computed as $\sqrt{41.467 / \big( 45 A^{-1/3} - 25 A^{-2/3} \big)}$, assuming the harmonic oscillator basis when evaluating the nuclear response functions.} The factor $\frac{m_T^2}{m_N^2}$ arises from the conventional relativistic normalization of states. The form of the differential cross section is given by \cite{Fitzpatrick:2012ix},
\begin{equation}
\label{eq:cross_section_DM-nucleus}
\frac{d \sigma_T^{NR}}{d \cos \theta} = \frac{1}{32 \pi (m_\chi + m_T)^2} \frac{1}{N_i}\sum_{i,j} \big| \mathcal{M}_{T}^{NR} \big|^2.
\end{equation}

Now, our task is to determine the coefficients associated with various nuclear operators. This involves approximating the relativistic high-energy model to its non-relativistic counterpart. In order to carry out this approximation, we must make certain assumptions regarding the low-energy regime of these interactions. Specifically, we assume that the operator scales as $\mathcal{O} (|\vec{p}_i|^2 / m_i^2)$ for any particle $i$ with mass $m_i$ and 3-momentum $\vec{p}_i$ involved in the interaction, including both the DM candidates and the nucleons. For more detailed derivations and approximations for bispinors, along with their results, we refer the reader to Ref.~\cite{Cirelli:2013ufw}.

\subsubsection{DM - nuclei interaction with a dark photon}
From the Lagrangian in \Cref{eq:Lagrangian}, which characterises interactions mediated by the dark photon, we deduce the following interaction term between DM particles $\chi$ and nucleons $N$, assuming $m_{Z^\prime}^2 \gg |q^2|$,
\begin{equation}
\label{eq:nuclei_nucleonH}
\begin{split}
\mathcal{H}_I^{Z^\prime N} = \frac{g_D g_{Z^\prime N}}{m_{Z^\prime}^2} \bigg[ g_V^\chi \overline{u}(p_3) \gamma_\mu u(p_1) \overline{u}(p_4) \gamma^\mu u(p_2) - g_A^\chi \overline{u}(p_3) \gamma_\mu \gamma^5 u(p_1) \overline{u}(p_4) \gamma^\mu u(p_2)  \bigg].
\end{split}
\end{equation}
Considering that the operator $\hat{v}^\perp \cdot \hat{v}^\perp$ yields no contribution, along with any other vector that is a linear combination of velocities due to the Galilean invariance of the non-relativistic Hamiltonian~\cite{Catena:2015uha}, the first term inside the square brackets is,
\begin{equation}
\label{eq:term1}
\begin{split}
\overline{u}(p_3) \gamma_\mu u(p_1) \overline{u}(p_4) \gamma^\mu u(p_2) \simeq & \ \ 4 m_\chi m_N + 2 m_\chi m_N \Bigg( i \hat{s}_N \cdot \bigg[ \frac{\hat{q}}{m_N} \times \hat{v}^\perp \bigg] \Bigg) + 2 m_N^2 \Bigg( i \hat{s}_\chi \cdot \bigg[ \frac{\hat{q}}{m_N} \times \hat{v}^\perp \bigg] \Bigg) 
\\& \ \ \ \ - \frac{m_\chi}{m_N} \vec{q}^2 - 4 \hat{q}^2 \hat{s}_\chi \cdot \hat{s}_N + 4 m_N^2 \bigg(\hat{s}_\chi \cdot \frac{\hat{q}}{m_N} \bigg) \bigg(\hat{s}_N \cdot \frac{\hat{q}}{m_N} \bigg)
\end{split}
\end{equation}
and the second term is,
\begin{equation}
\label{eq:term2}
\begin{split}
\overline{u}(p_3) \gamma_\mu \gamma^5 u(p_1) \overline{u}(p_4) \gamma^\mu u(p_2) \simeq & \ \ 8 m_\chi m_N \Bigg( \hat{s}_\chi \cdot \hat{v}^\perp + i \hat{s}_\chi \cdot \bigg[ \hat{s}_N \times \frac{\hat{q}}{m_N} \bigg] \Bigg).
\end{split}
\end{equation}
Notice that this expression is general, indicating the interaction of DM with  nuclei of any type. The terms proportional to $\vec{q}^{\,2}$ are not proportional to any of the masses, and since there is a factor of $m_{Z^\prime}^2$ in the denominator of \Cref{eq:nuclei_nucleonH}, we can neglect its contribution. In this way, the interaction term can be written in the non-relativistic basis of \cite{Catena:2015uha} as,
\begin{equation}
\label{eq:nuclei_NRH}
\begin{split}
\mathcal{H}_{I, NR}^{Z^\prime N} 
= c_1^N \mathcal{O}_1^{NR} + c_3^N \mathcal{O}_3^{NR} + c_5^N \mathcal{O}_5^{NR} + c_6^N \mathcal{O}_6^{NR} + c_8^N \mathcal{O}_8^{NR} + c_9^N \mathcal{O}_9^{NR}
\end{split}
\end{equation}
where $c_i^N$ denotes each prefactor in \Cref{eq:term1} and \eqref{eq:term2}. Next, to determine $g_{Z^\prime N}$ from the interaction with quarks, we observe that the interaction is purely vectorial in nature, as noted in~\cite{Cirelli:2013ufw}: $g_{Z^\prime p} = 2 g_{Z^\prime u} + g_{Z^\prime d}$ and $g_{Z^\prime n} = g_{Z^\prime u} + 2 g_{Z^\prime d}$.

From here, we can proceed to compute the differential cross section as outlined in \Cref{eq:scattering_amp}. However, analytical results are not presented here, as all calculations have been performed numerically.

\subsubsection{DM - nuclei interaction with a scalar}

From the scalar interactions described in \Cref{eq:Lagrangian}, we derive the following interaction term between DM particles $\chi$ and nucleons $N$, under the assumption that $m_{\Phi}^2 \gg |q^2|$:
\begin{equation}
\label{eq:nuclei_nucleonS}
\begin{split}
\mathcal{H}_I^{\Phi N} = \frac{g_D g_{\Phi N}}{m_{\Phi}^2} \bigg[ \overline{u}(p_3)u(p_1) \overline{u}(p_4) u(p_2) \bigg].
\end{split}
\end{equation}
Applying the same procedure as with the dark photon, we determine the non-relativistic interaction term between a nucleon $N$ and the scalar $\Phi$. This is,
\begin{equation}
\label{eq:nuclei_NRScalar}
\begin{split}
\mathcal{H}_{I, NR}^{\Phi N} 
&\equiv 4 g_D g_{\Phi N} \frac{m_\chi m_N}{m_{\Phi}^2} \mathcal{O}_1^{NR}.
\end{split}
\end{equation}
To express $g_{\Phi N}$ in terms of the couplings to the quarks, we follow the approach outlined in \cite{Cirelli:2013ufw}, considering the absence of direct coupling to gluons at tree level, thus:

\begin{equation}
\begin{split}
g_{\Phi N} = \sum_{q=u,d,s} \frac{g_{\Phi q}}{m_q} m_N f_{Tq}^{(N)} + \frac{2}{27} f_{TG}^{(N)} \sum_{q=c,b,t} \frac{g_{\Phi q}}{m_q} m_N
\end{split}
\end{equation}
where $f_{Tq}^{(n)} = 0.0110, 0.0273, 0.0447$ and $f_{Tq}^{(p)} = 0.0153, 0.0191, 0.0447$ for $q= u,d,s$ respectively; $f_{TG}^{(N)} = 0.917$, as can be found in \cite{Bell:2021fye}.

\section{Results and discussion}
\label{sec:results}
\begin{figure}
    \centering
    \includegraphics[width=0.49\linewidth, draft=false, clip=true]{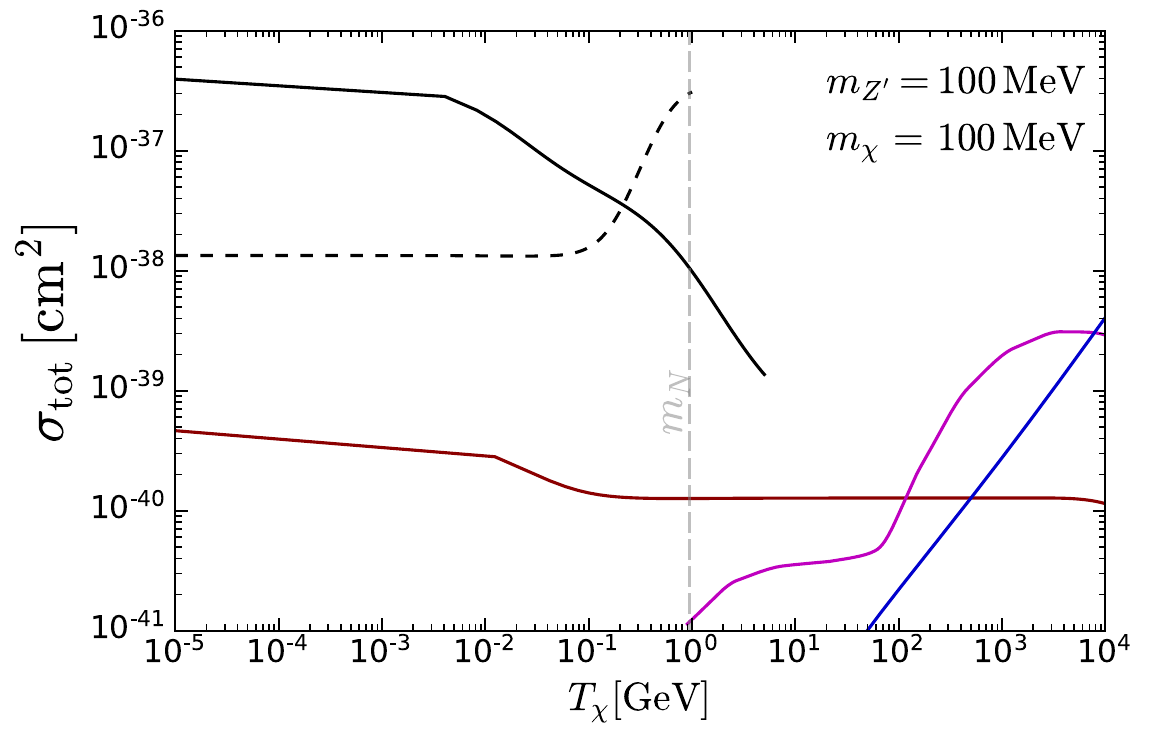}
    \includegraphics[width=0.49\linewidth, draft=false, clip=true]{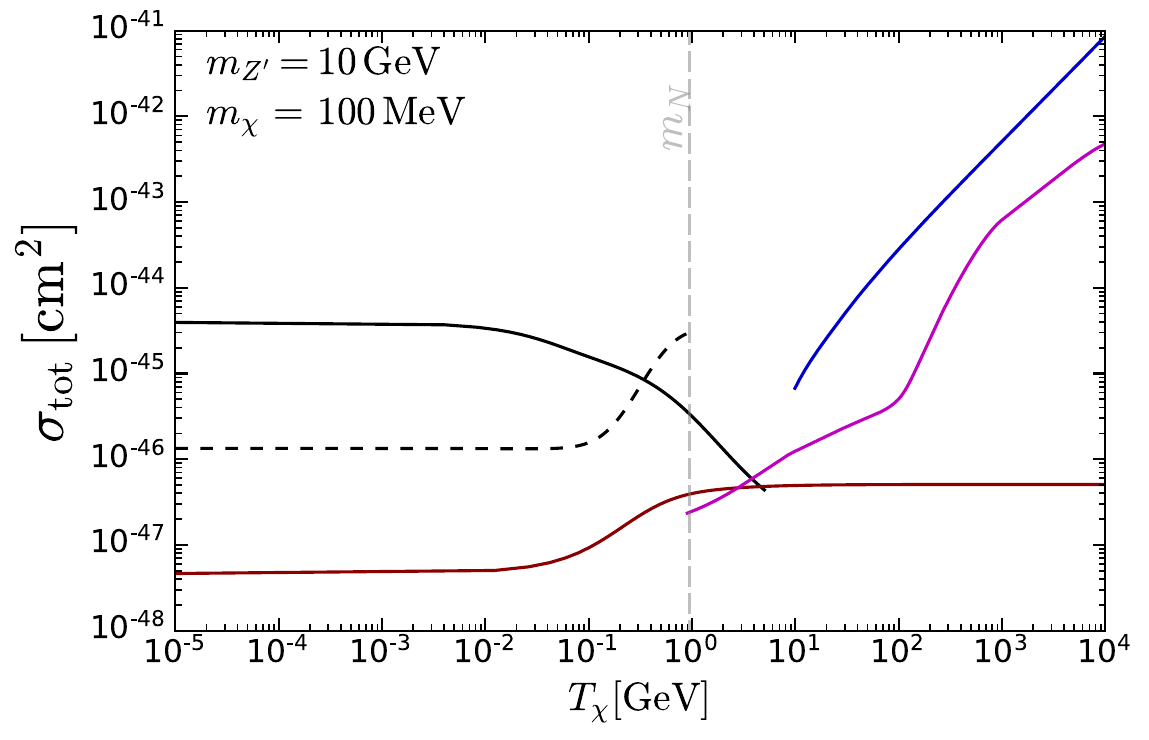}
    \includegraphics[width=0.7\linewidth, draft=false, clip=true]{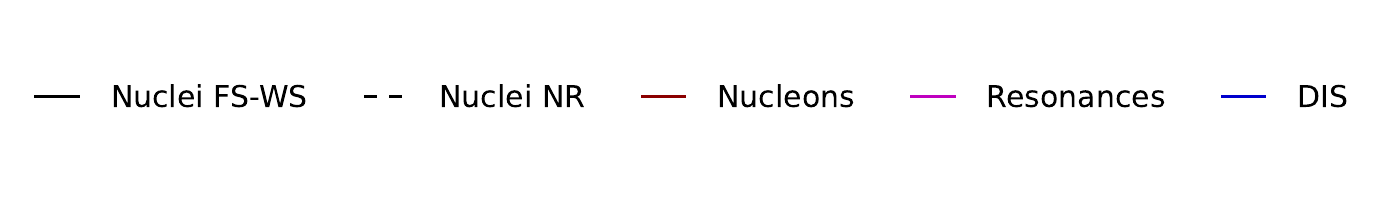}
    \caption{DM interaction cross section mediated by a dark photon as a function of kinetic energy, shown for various benchmark values of DM and dark photon masses. The kinetic mixing with the SM is set to $\epsilon = 10^{-5}$, and the dark photon coupling to the dark sector is fixed at $g_D = 0.1$. The plot explores four distinct interaction regimes, represented by different colours: the blue line indicates DIS, the magenta line corresponds to resonant scattering, the brown line shows interactions with nucleons, and the black line represents interactions with nuclei. For the latter, the solid and dashed lines differentiate the two approaches used to describe the interactions with nuclei.
 }
    \label{fig:Vsigma}
\end{figure}

In this section, we analyse the DM scattering cross section and capture rate for different energy regimes, focusing on the influence of varying DM and dark photon (scalar) masses. For our study, we use a WD with a of mass $M_\star=1\,\rm M_\odot$ and a radius of  $R_\star= 5.7\times10^3\,\rm km$. The radial profiles for this WD have been computed by applying the Tolman-Oppenheimer-Volkoff (TOV) equations~\cite{PhysRev.55.364,PhysRev.55.374}, describing the hydrostatic equilibrium of a spherically symmetric, non-rotating star, coupled to the Salpeter equation of state (EOS)~\cite{salpeter61}.

\begin{figure}
    \centering
    \includegraphics[width=0.5\linewidth, draft=false, clip=true]{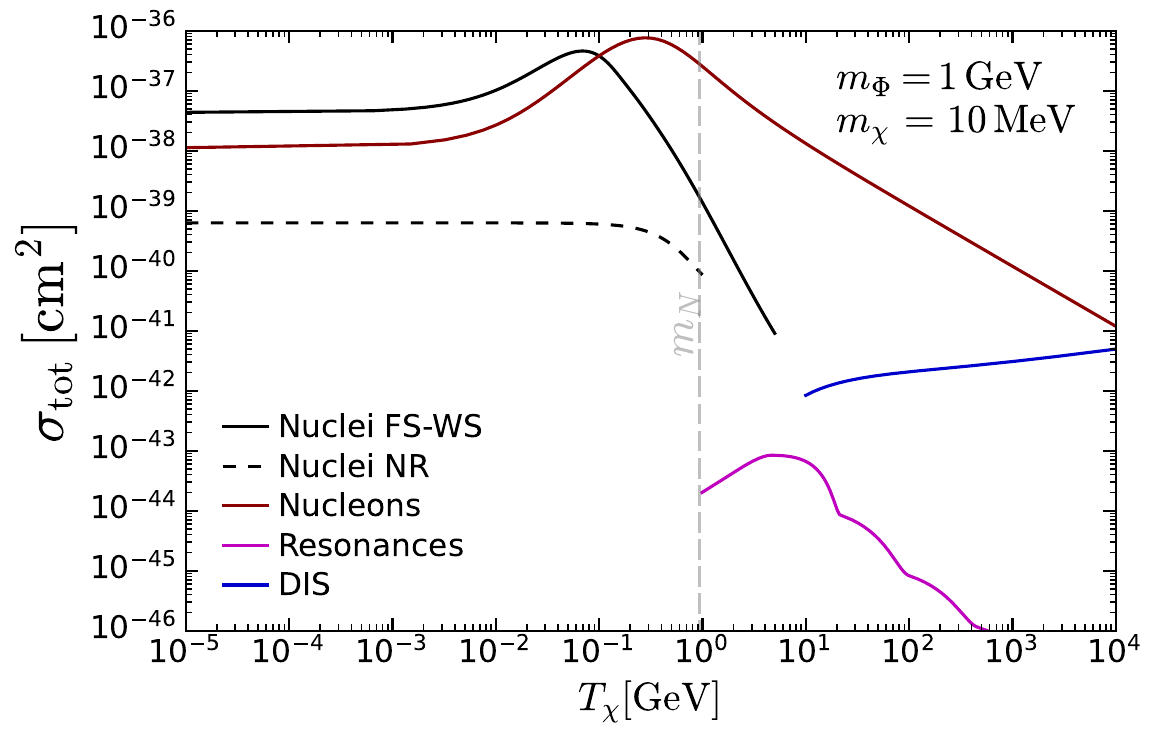}
    \caption{DM interaction cross section mediated by a dark scalar as a function of kinetic energy, shown for various benchmark values of DM and dark scalar masses. The couplings are set to $g_{N\Phi} = 10^{-5}$ with the SM and $g_D = 0.1$ in the dark sector. As in~\Cref{fig:Vsigma} for the dark photon case, the plot explores four interaction regimes: DIS (blue), resonant scattering (magenta), nucleon interactions (brown), and nuclear interactions (black). Solid and dashed lines in the black curve distinguish the two approaches for nuclei interactions.}
    \label{fig:Ssigma}
\end{figure}

The analysis provides insights into the dominance of different interaction regimes—such as DIS, resonant inelastic scattering, and scattering with nucleons and nuclei—under different conditions. To explore the DM scattering cross section and capture rate across different energy regimes, we have selected two benchmark points for the vector and one for the scalar. For the vector, we choose the couplings to be proportional to the electromagnetic ones, $g_{Z^\prime q} \equiv \epsilon e Q_q$, where $e$ is the electromagnetic coupling and $Q_q$ is the electromagnetic charge of quark $q$. This choice corresponds to cases where the connection between the dark photon and the SM is established through kinetic mixing~\cite{Ballett:2019pyw, Abdullahi:2020nyr, Abdullahi:2023tyk}. We fixed the mass of $\chi$ to $100$ MeV and explored a light ($100$ MeV) and heavy ($10$ GeV) mediator cases. The choice of $m_\chi$ was based on realistic values to have highly boosted DM, see for instance the values taken by \cite{Wang:2021jic} or \cite{Bringmann:2018cvk} for boosting DM. For the scalar, we chose a rather lower mass of $\chi$, $10$ MeV, and explored just the case of $m_\phi = 1$ GeV. The choice for the scalar was just based on the different behaviour of the nucleon interaction, as seen on \Cref{fig:Ssigma}. We also set the vector coupling with the SM  to $10^{-5} e Q_q$ ($g_{N\Phi}= 10^{-5}$), and the dark photon (scalar) coupling to the dark sector to $g_D = 0.1$. \response{We have chosen DM masses we are less sensitive to in direct detection experiments\footnote{See for instance Figures (5.4) and (5.5) in \cite{Cirelli:2024ssz}.}, i.e., sub-GeV DM. In the case of the mediator masses, we are interested in also showing sub-GeV cases, since these have been widely studied in what is called ´´dark sectors"~\cite{Ballett:2019pyw}. Particularly, we just selected one benchmark for the scalar, since the different masses exhibited a similar behaviour. For the couplings, because they just affect the overall normalization, we chose numbers that put us away from the geometric limit, which would require a different treatment of the capture rates~\cite{Bell:2020jou}.} We set all scalar quark coefficients to $g_q^S = 1$ and assume the dark photon couplings to the DM particles are purely vectorial. The couplings selection was aimed to be far from the regime where an analysis of the opacity of the star is needed, which is out of the scope of the present analysis.

In \Cref{fig:Vsigma}, we present the cross section as a function of the kinetic energy of the incoming DM far from the star for scenarios where the DM interaction with stellar matter is mediated by a dark photon. The four different regimes are illustrated as follows: blue for DIS, magenta for resonance, brown for scattering with nucleons, and black for scattering with nuclei. Note that for the latter case, we employed two different approaches to describe the effective interaction of DM with the nucleus. The black dashed lines represent the non-relativistic (NR) approach, as detailed in \cref{sec:nuclei}, while the black solid line uses the Fermi-Symmetrized Woods-Saxon (WS) form factor, presented in \Cref{Ap:Helm_FF_computation}. There are two important remarks. First, the NR approach is more reliable for interactions with nuclei at low kinetic energies, as it provides a more comprehensive description of the nucleus by incorporating the nuclear shell model and energy level transitions. However, this approach is limited to non-relativistic conditions. For higher kinetic energies, the WS form factors are more suitable, as they do not require a complete non-relativistic account of the detailed nuclear shell structure. We consider using the NR approach when $T_\chi < 0.5 \times m_\mathrm{min}$, where $T_\chi$ is the kinetic energy of the incoming DM particle at the moment of scattering, and $m_\mathrm{min}$ is the smallest mass involved in the interaction, which will always be that of the DM particle in our cases. Secondly, for inelastic interactions it is important to note that the resonant inelastic  and DIS interactions  overlap in phase space. We should not simply add  them up, as this would lead to overcounting. The DIS cross section should be computed only for $W \gtrsim 1.8\,\rm GeV$ to avoid overcounting contributions from overlapping inelastic scattering through resonances. For the cases we are dealing with, this overlap region has negligible contribution from the DIS cross section. For computing it, we have used the parton library in Python,\footnote{See https://github.com/DavidMStraub/parton.} which is based in the LHAPDF project~\cite{Buckley:2014ana}.

In the left panel, where the DM particle mass is $m_\chi = 100\,\rm MeV$ and the dark photon mass is $m_{Z^\prime} = 100\,\rm MeV$, in the high-energy regime the resonance dominates over DIS.  As the kinetic energy approaches the nucleon mass, interactions with nucleons become predominant. Observe that in the region where these interactions dominate, the DM cross section does not depend on the kinetic energy and it stabilises on the limit $T_\chi \to 0$. At lower kinetic energies ($T_\chi << m_N$), scattering with nuclei becomes more significant. Here, the two approaches show markedly different behaviours, differing by more than one order of magnitude and coinciding only around $T_\chi \sim \mathcal{O}(100),\rm MeV$. The use of FS-WS form factors might be overestimating the cross section with nuclei at very low energies, where the nuclear structure becomes more important. On the other hand, when the energies become less non-relativistic, the NR model starts to deviate from the lower values predicted, since the computation relies on high masses with respect to the momenta. A mixed approach is needed not to over- or underestimate the cross sections.

The behaviour in the low regime (nuclei and nucleon) is consistent across all benchmarks for DM and dark photon masses. The primary difference appears in the high-energy regime. 
For example, in the right panel, where the DM mass remains the same but the dark photon mass is $m_{Z^\prime} = 10\, \rm GeV$, DIS  dominates across the energy range compared to the resonance case. However, the resonant cross section still competes with DIS, and even with nucleon interactions, within a small region around $\mathcal{O}(1) - \mathcal{O}(10)\, \rm GeV$.

In \Cref{fig:Ssigma}, we present the case where DM interactions are mediated by a dark scalar. 
As in the vector mediator case (\Cref{fig:Vsigma}), we explore four energy regimes: DIS, resonant scattering, nucleon interactions, and nuclear interactions. 
Notice that the same two approaches to describe interactions with nuclei, represented by solid and dashed lines, are also applied here.
For this scenario, we have fixed the dark scalar mass to $m_{\Phi}=1\,\rm GeV$ while taking two DM mass as $m_\chi= 10 \,\rm MeV$.
The low-energy regime is relatively similar to the case of DM interactions mediated by vectors. However, the behaviour of DM interactions with nucleons (brown line) is different. Contrary to the dark photon case, where the cross section with nucleons is relatively constant, here the cross section strongly depends on the DM kinetic energy.
This behaviour can be understood from \Cref{eq:DM_N_Sxsec}, where the total energies of the DM ($E_1$) and the nucleon ($E_2$) compete in the denominator. When $E_1$ becomes smaller than $E_2$, the cross section becomes nearly constant below the nucleon mass (around $1$ GeV). In contrast, at higher DM energies, the cross section is suppressed as $E_1^{-2}$, due to the dominance of $E_1$ in the denominator. Finally, the high-energy regime is dominated by DIS. We observe that the DIS cross section is suppressed by several orders of magnitude compared to those of nucleons and nuclei, with resonant scattering being even more suppressed. The latter presents a maximum around $T\chi \sim 5$ GeV related to the scalar dipole of the form factors. We present only one benchmark, as all scenarios exhibit similar behaviour regardless of variations in the dark scalar and DM masses.

%
\begin{figure}
\centering
    \includegraphics[width=0.49\linewidth, draft=false, clip=true]{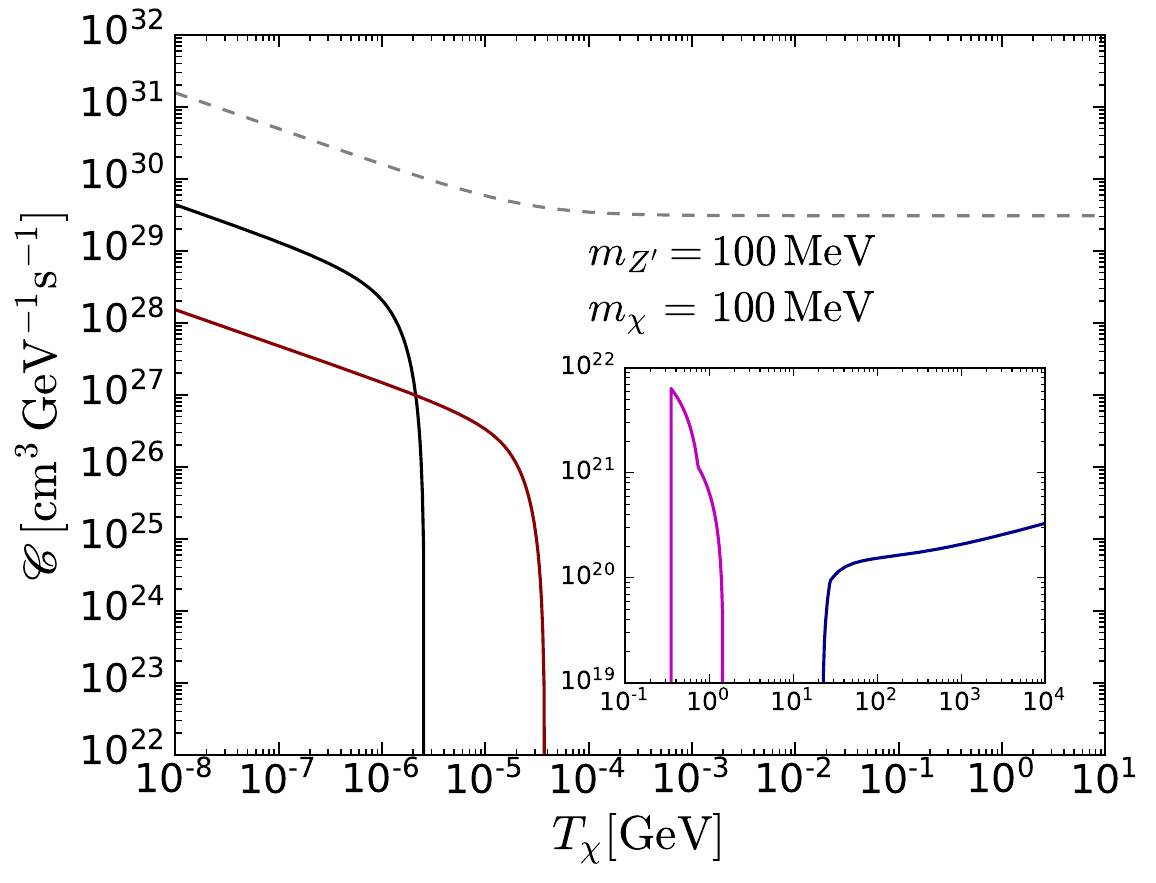}
    \includegraphics[width=0.49\linewidth, draft=false, clip=true]{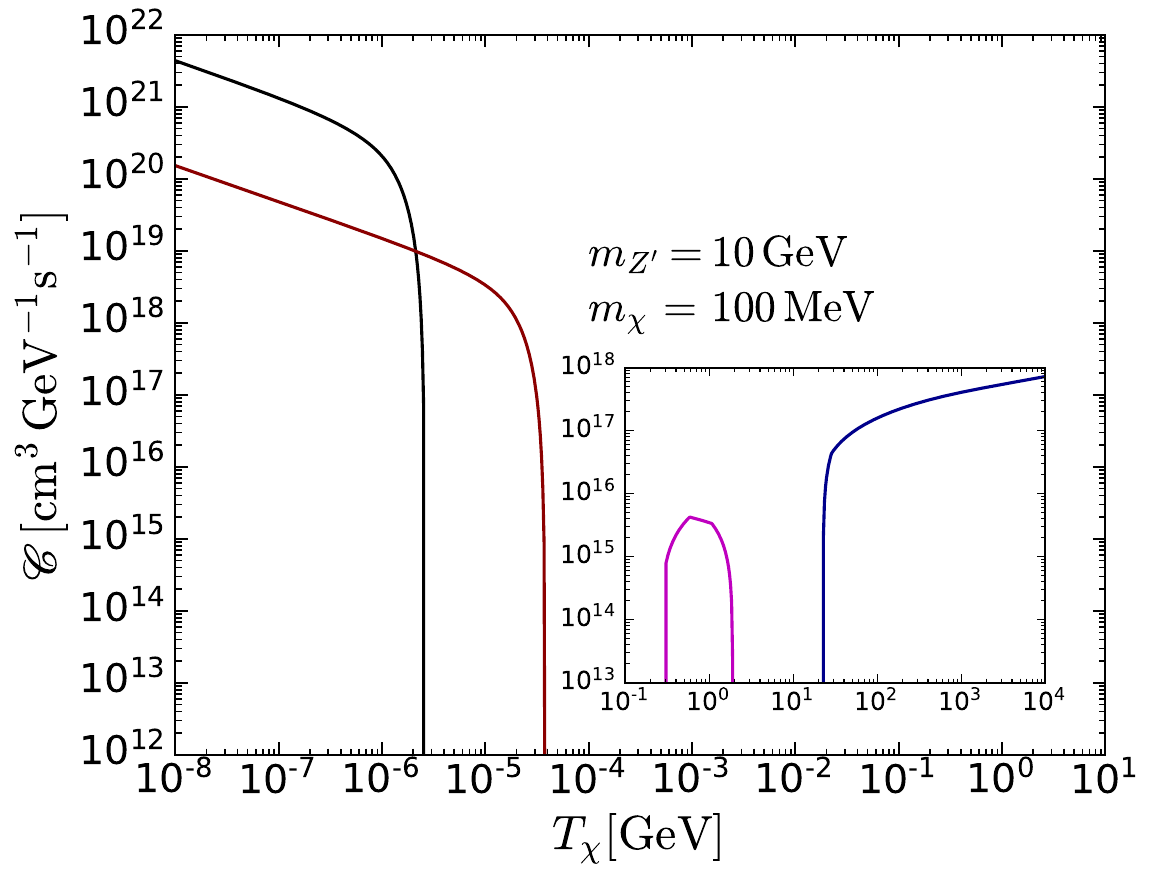}
    \includegraphics[width=0.7\linewidth, draft=false, clip=true]{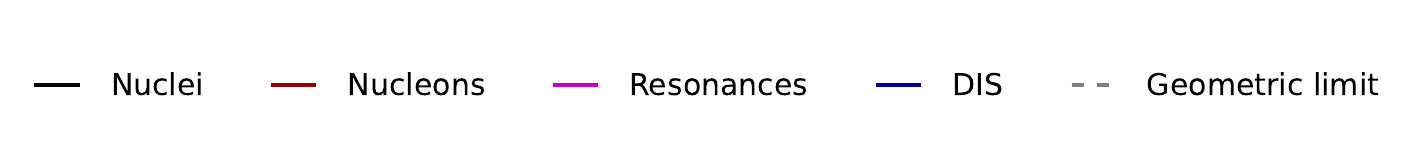}
    \caption{DM capture rate density as a function of DM kinetic energy for the case where DM interactions are mediated by a dark photon. The benchmark values of DM and dark photon mass are taken as $m_Z=0.1, 10\, \rm GeV $ and $m_\chi =100\, \rm MeV$. The kinetic mixing with the SM is set to $\epsilon = 10^{-5}$, and the dark photon coupling to the dark sector is fixed at $g_D = 0.1$. The plot explores four distinct interaction regimes, represented by different colours: the blue line indicates DIS, the magenta line corresponds to resonant scattering, the brown line shows interactions with nucleons, and the black line represents interactions with nuclei. The grey dashed line shows the geometric limit.}
    \label{fig:V_capture_rate}
\end{figure}
%

To explore the DM density capture rate in the scenario where DM interactions are mediated by a dark photon, we use the two benchmark points from \Cref{fig:Vsigma}. The choice of these masses is motivated by the cross section behaviour, where resonant scattering dominates for lighter mediators, while DIS becomes the dominant process for heavier mediators. For the chosen couplings, the capture rate density stays within the optically thin limit, not exceeding the maximum capture rate (gray dashed line).
The capture rate density is presented for all energy regimes. As expected, in the low-kinetic energy region, the dominant processes are scattering with nuclei (black) and nucleons (brown). In this regime, the capture density rate is larger for lighter mediators, as it is proportional to the scattering cross section, which is inversely proportional to the squared mediator mass ($\propto m_{Z^\prime}^{-2}$). Therefore, heavier mediators result in smaller cross sections and a suppressed capture density rate.

As the kinetic energy increases, a gap appears in the range $T_\chi \sim \mathcal{O}(10^{-4}-10^{-1})\,\rm GeV$, indicating that no capture process is viable within this energy range. This occurs because, in elastic interactions, after a certain threshold outgoing DM particles with a velocity less than the escape velocity are out of the phase space. The stronger kinematic constraints of the elastic process ($2 \to 2$) contribute to this. At higher energies, inelastic processes, such as resonances ($2 \to 3$) or DIS ($2 \to N$), allow slower outgoing DM particles within the permitted phase space. Around kinetic energies of $1\,\rm GeV$, there is a small probability of capture occurring via resonant scattering (magenta), though the energy window for this process is quite narrow. In the high-kinetic energy regime, DIS (dark blue) becomes dominant, as expected, because the production of jets increases the available phase space, allowing for more damped outgoing DM. While the capture probability is small, the energy range for DIS is broader than that for resonant scattering. Notably, capture via resonant scattering is more efficient than DIS when the dark photon is lighter (left panel). However, for smaller cross sections and larger dark photon masses (right panel), DIS dominates over resonant scattering in terms of capture probability.

%
\begin{figure}
\centering
    \includegraphics[width=0.5\linewidth, draft=false, clip=true]{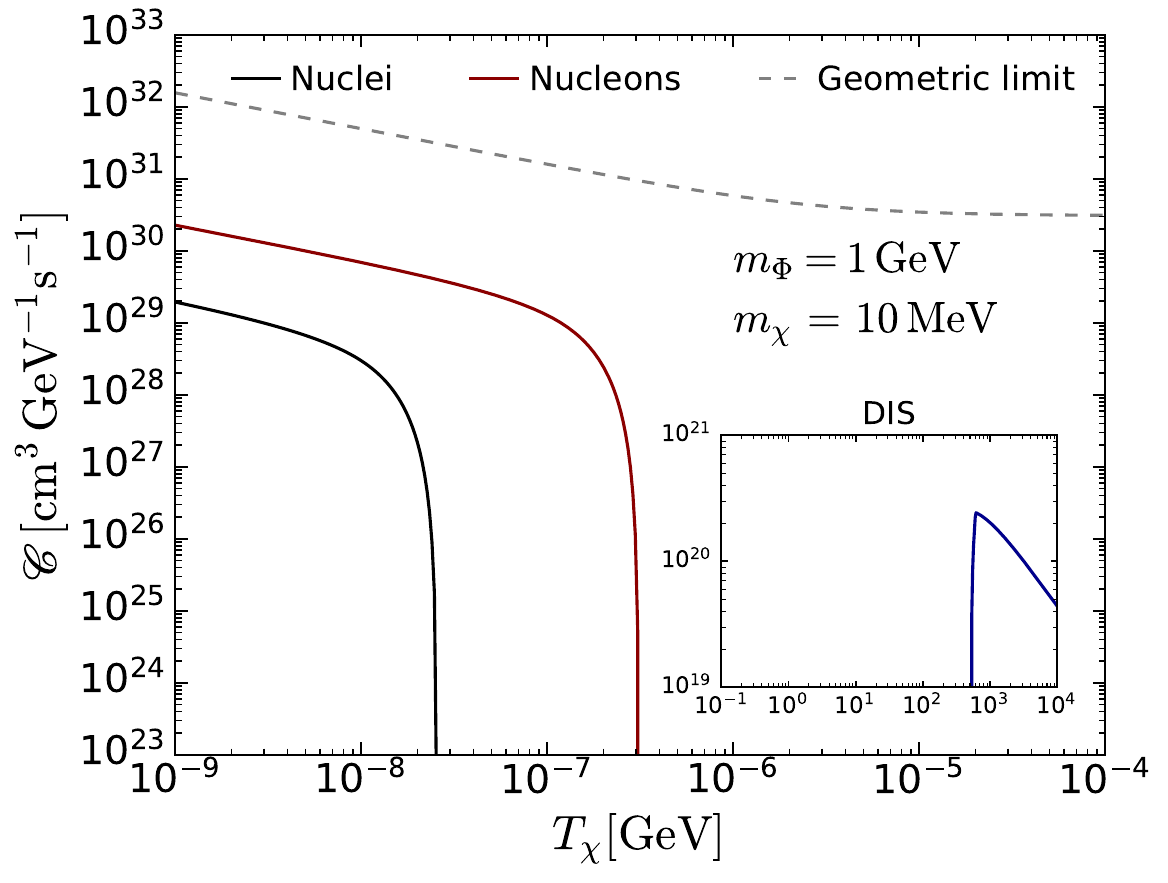}
    \caption{DM capture rate density as a function of DM kinetic energy for the case where DM interactions are mediated by a dark scalar. The benchmark values of DM and dark photon mass are taken as $m_\Phi=1 \, \rm GeV $ and $m_\chi =10\, \rm MeV$. The couplings are set to $g_{N\Phi} = 10^{-5}$ with the SM and $g_D = 0.1$ in the dark sector. As in~\Cref{fig:V_capture_rate}, the plot shows different interaction regimes: DIS (blue), nucleon interactions (brown), and nuclear interactions (black). The grey dashed line shows the geometric limit.}
     \label{fig:S_capture_rate}
\end{figure}
%

We also explore the DM capture rate density when DM interactions are mediated by a dark scalar. Since the DM cross section is similar across all benchmarks in this scenario, in \Cref{fig:S_capture_rate}, we present only the case where the dark scalar mass is $m_\Phi = 1\,\rm{GeV}$ and the DM mass is $m_\chi = 10\,\rm{MeV}$. In this case, we set the coupling of the dark scalar to the SM to $g_{N\Phi} = 10^{-5}$ and $g_D = 0.1$. The results for the capture density rates exhibit the same overall behaviour as for the dark photon mediator; however, we observe that there is no possibility of capture occurring via resonant scattering, which has no contribution for sufficiently slow DM particles that can be captured. As a result, the range of energies where DM capture is not possible is extended, covering $T_\chi \sim \mathcal{O}(10^{-7} - 10^{2})\,\rm{GeV}$. 
Above these energies, the capture rate density is extremely suppressed in the narrow window of energies (above $500$ GeV) in which capture by DIS would be, in principle, possible. 
This suggests that, for this DM candidate, the effects on WDs at these energies are unlikely to be noticeable. 

\response{Finally, it is important to consider how the effects of DM capture in WDs could be detected. Although we are currently far from the required sensitivity, DM can deposit kinetic energy during the capture process and potentially through subsequent decay or annihilation into SM particles, leading to additional heating of old and cold WDs. By observing their luminosities, one could, in principle, constrain the DM parameter space. This possibility has been widely discussed in the literature~\cite{Bertone:2007ae, McCullough:2010ai, Panotopoulos:2020kuo, Garani:2023esk, Leane:2023woh, Leane:2024bvh, Bramante:2023djs}, though not specifically in the context of boosted DM.}

The results presented here for both scenarios, vector- and scalar-mediator, highlight the challenges in detecting high-energy DM interactions, particularly for scalar mediators where capture at these energies is very challenging. A natural next step in refining these estimates involves, for example, considering  DM boosted by  cosmic ray interactions~\cite{Bringmann:2018cvk} or by blazars~\cite{Wang:2021jic}. This would provide a more realistic BDM flux in regions of the universe where the DM density is well understood, such as our galaxy. By offering a more accurate characterisation of DM fluxes, this approach potentially enable us the establishment of constraints on DM interactions with WD components in the sub-GeV mass regime, especially for DM interactions that depend on velocity and/or transferred momentum. Such interactions cannot be effectively probed in the context of non-boosted local DM, as it is non-relativistic. While this is beyond the scope of the current work, it represents an important step forward, which we leave open for future work.

\section{Conclusion}
\label{sec:conclusion}

In this work, we have described the capture of DM particles model in WDs across a full energy regime, from very low energies to higher energies. Our approach differs from conventional DM capture in two main ways. 
First, we have introduced the DM flux as a delta function to analyze the effect of specific energies on the capture rate of a WD, thereby enabling to account for incoming DM from different sources. This has been achieved by assuming that DM is isotropically boosted from various energies throughout the Universe, so that the flux arrives at the star from all directions. This approximation is applicable even in the non-relativistic case, as we are considering a broader perspective on DM rather than focusing solely on the local population typically modelled by a Boltzmann distribution.  While this approach yields a rough estimation of the capture probability, it is essential for understanding the broader implications of DM interactions. Second, we have detailed how the energy range affects the interactions between DM and stellar matter. We assume a fermionic DM candidate interacting with stellar matter through either a dark photon or a dark scalar. At high energies, interactions may proceed via mechanisms such as deep inelastic scattering or inelastic scattering through the production of $N-$ and $\Delta-$ resonances, while at lower energies, interactions with nucleons or, in the non-relativistic regime, with nuclei become more relevant. In particular, inelastic interactions involving the production of resonances were computed for the first time for DM interacting with a vector or scalar mediator, using the Feynman-Kislinger-Ravndal approach along with the Rein and Sehgal treatment of the problem.

We presented the DM cross section for interactions mediated by either a dark photon or a dark scalar across a wide range of energies $(T_\chi \in [10^{-5}, 10^4]\,\rm GeV)$. For the dark photon case, we selected two benchmark points, fixing the DM mass to $m_\chi = 100\,\rm MeV$ and the dark photon mass to $m_{Z^\prime} = 100\,\rm MeV$ and $1\,\rm GeV$. In both cases, the non-relativistic behaviour remains consistent, with the main differences arising in the regime of high kinetic energy. This regime dominates over direct interaction with DM for dark photon masses of the order of the DM mass. Conversely, if the dark photon mass is heavier, we observe that DIS and resonant interactions become very similar, with DIS dominating  the kinetic energy range and resonant scattering only around $T_\chi\sim10^3\,\rm GeV$. 
On the other hand, the scalar case remains consistent regardless of the choice of masses. Therefore, we only present one benchmark point for this scenario. In contrast to the dark photon case, the interactions with nuclei are suppressed as $E_2^{-1}$. Moreover, the resonant scattering cross section is suppressed compared to the DIS interaction at the energies considered. 

We have also presented the capture rate density. Here, the kinetic energy range is even broader because it corresponds to the DM particle's energy far from the star. As a result, the energies considered for the interactions are significantly higher due to the gravitational acceleration of the particles. For the vector mediator case, we have discovered that, although the contributions coming from low energy nuclei and nucleons interactions are higher, DIS and resonant interactions can also mediate the capture of DM for high energy incoming particles. However, there is a gap in energies, $T_\chi \sim \mathcal{O}(10^{-4}-10^{-1})\,\rm GeV$, for which no capture is possible and the WD is blind to incoming DM. For the case of a scalar mediator, capture of high energy incoming particles is very suppressed and just possible for DIS.

Finally, future improvements could focus on incorporating realistic contributions from BDM, such as  DM boosted by cosmic rays or blazars. This could improve the framework by enabling a more accurate characterisation of DM fluxes and making it possible to set constraints on DM interactions with WD components. While this is beyond the scope of this work, it opens a promising avenue for future studies.

\acknowledgments
We want to thank professors Paolo Finelli and Nicol\`o Masi for useful discussions about interactions with nuclei, Dr. Marvin Ascencio for the references and discussions on inelastic interactions through the production of resonances and Dr. Alessandro Granelli for his suggestions about interactions with nucleons through scalars and the DIS framework.  We would also like to thank Suroor Seher Gandhi for engaging in discussions related to this work.

This research  has received support from the European Union’s Horizon 2020 research and innovation programme under the Marie Sk\l{}odowska-Curie grant  agreement No 860881-HIDDeN (JHZ), 
JSPS KAKENHI: 22KJ1022 (SH) and
the Cluster of Excellence “Precision Physics, Fundamental Interactions, and Structure
of Matter” (PRISMA+ EXC 2118/1) funded by the Deutsche Forschungsgemeinschaft (DFG, German Research Foundation) within the German Excellence Strategy (Project No. 390831469) (MRQ). JHZ is also supported by the National Science Centre, Poland (research grant No. 2021/42/E/ST2/00031).

\appendix
%
\section{DM velocity relations}
\label{app:DM_velocities}
Considering the metric
\begin{equation}
d_s^2 =
   \begin{cases}
         - g_{tt} \,dt^2+ g_{rr}\, dr^2+r^2(d\theta^2+\sin^2\theta d\phi^2), & \text{at } r= R_\star\\
          & \\
         - dt^2+dr^2+r^2(d\theta^2+\sin^2\theta d\phi^2), & \text{at } r= \infty\\
   \end{cases}
\end{equation}
we derive the relationship between DM velocities near and far from the star, as given in~\Cref{eq:new_w_u_relation}. To establish this relationship, we use the conservation of energy $E = -g_{\mu\nu} p^\mu \xi^\nu$, where $\xi = (1, 0, 0, 0)$. For the two different scenarios, (I) when the DM has a velocity $u_\chi$ far from the star (at $\infty$), and (II) when the DM has a velocity $\omega$ near the WD, the energy is given in both cases by 
\begin{align}
        E_{\rm I} &= \frac{m_\chi}{\sqrt{1-u^2_\chi}}\\
        E_{\rm II}&=  g_{tt} m_\chi u_{\rm II}^t,
\end{align}
where $u_\chi^t = 1/(A\, \sqrt{1-u^2_\chi})$.  From here, we can determine the radial velocity $u^r_{\rm II}$ near the WD. Using the relation
\begin{equation}
    g_{\mu\nu} u^\mu_{\rm II}u^\nu_{\rm II} = -1
\end{equation}
and considering only radial motion, we have
\begin{align}
    g_{tt}u^t_{\rm II} u^t_{\rm II}+g_{rr}u^r_{\rm II}u^r_{\rm II}=&-1\\[0.2cm]
    A (u^t_{\rm II})^2+B(u^r_{\rm II})^2 =& -1,
\end{align}
where $A= g_{tt}$ and $B= g_{rr}$. Solving for $u^r_{\rm II}$ and using the definition of $u^t_{\rm II}$, we obtain
\begin{equation}
\label{eq:velocities_II}
    u_{\rm II}^r =\sqrt{\frac{1}{A\,B\,(1-u_\chi^2)}-\frac{1}{B}}\,\,\ \to\ \,\,\frac{u^r_{\rm II}}{u^t_{\rm II}}\equiv\frac{dr}{dt}= A\,\sqrt{\frac{1}{A\,B}-\frac{\,(1-u_\chi^2)}{B}}.
\end{equation}
Next, we move to local variables using the transformation matrices $e^{\hat{r}}_r$ and $e^{\hat{t}}t$. These can be found from the conditions $g_{\hat{t}\hat{t}} = -1$ and $g_{\hat{r}\hat{r}} = 1$, which implies $e^{\hat{t}}_t = \sqrt{A}$ and $e^{\hat{r}}_r = \sqrt{B}$. The velocity of the DM near the WD, i.e., in the hatted frame, is then 
\begin{equation}
    \omega=\frac{d\hat{r}}{d\hat{t}}=\frac{e^{\hat{r}}_r}{e^{\hat{t}}_t}\frac{u^r_{\rm II}}{u^t_{\rm II}},
\end{equation}
and squaring, we finally arrive at the expression given in~\Cref{eq:new_w_u_relation}.

\section{Resonant scattering with dark matter}
\label{sec:App_resonances}
In this appendix, we provide the details for the computation of the resonant amplitude calculation. In order to perform all the computations, it is important to use the resonant or isobaric frame, as can be found in \cite{Kabirnezhad:2024cor}. In the case of the interaction: $\chi (p_1) N(p_2) \to \chi (p_3) N^{*}(p_4)$, the kinematical quantities in this frame in terms of the parameters from the Lab frame are:
\begin{align}
p_1& = (((E_1 - E_3)^2 + 2 m_N E_1 - Q^2) / (2W), A_{13}, 0, B_{13}^-),\\[0.1cm]
p_2& = (m_N(E_1 - E_3 + m_N) / W, 0, 0, - m_N Q / W),\\[0.1cm]
p_3& = ((-(E_1 - E_3)^2 + 2 m_N E_3 + Q^2) / (2W), A_{13}, 0, B_{13}^+),\\[0.1cm]
p_4 &= (0, 0, 0, W),\\[0.1cm]
A_{13}& = \sqrt{(p_1 + p_3 - Q)(p_1 - p_3 + Q)(-p_1 + p_3 + Q)(p_1 + p_3 + Q)}  / (2Q), \\[0.1cm]
B_{13}^{\pm}& = ((E_1^2 - E_3^2)(E_1 - E_3 + m_N) - (E_1 + E_3 \pm m_N) Q^2) / (2 W Q).
\end{align}
To proceed with the computation of the cross sections, we treat each case, vector or scalar, separately. However, the integration limits remain identical for all cases and are given by:
\begin{align}
   &m_N + m_\pi \le \  W \  \le \sqrt{s} - m_\chi,\\[0.1cm]
& q^2 = 2m_\chi^2 - \frac{\left(E_1 m_N + m_\chi^2\right) \left( s - W^2 + m_\chi^2 \right)}{s} \nonumber\\
&\,\,\,\,\,\,\,\,\,\,\,\,\,\,\,\,\,\,\,\,\,\,\,\,\,\,\,\,\mp \frac{\sqrt{\left(E_1 m_N + m_\chi^2\right)^2 - s m_\chi^2} \sqrt{\left( s - W^2 + m_\chi^2 \right)^2 - 4 s m_\chi^2}}{s},
%
%
\end{align}
where $\mp$ refers to $q^2_\mathrm{min}$ and $q^2_\mathrm{max}$, respectively.

\subsection{Vector Current and Coefficients for  resonant scattering}
\label{sec:App_resonances_vector}

The amplitude in the case of a photon mediator is expressed as:
\begin{equation}
\begin{split}
& \qquad \mathcal{M} (\chi(p_1, \lambda_1) N(p_2) \to \chi(p_3, \lambda_2) N^{*} (p_4)) \\
&= \frac{g_D g_{Z^\prime N}}{q^2 - m^2_{Z^\prime}} \left[ \overline{u}_{p_3 \lambda_2} \gamma_\mu \left( g_V - g_A \gamma^5 \right) u_{p_1 \lambda_1} \right] \left( g^{\mu \nu} - q^\mu q^\nu / m^2_{Z^\prime}  \right) \bra{N^{*}} J_\nu^{+} (0) \ket{N} \\
&= 2M \frac{g_D g_{Z^\prime N}}{q^2 - m^2_{Z^\prime}} V_{\lambda_1 \lambda_2}^\mu \bra{N^{*}} F_\mu^{V} \ket{N} \\
&= 2M \frac{g_D g_{Z^\prime N}}{q^2 - m^2_{Z^\prime}} \left( C_L^{\lambda_1 \lambda_2} e_L^\mu + C_R^{\lambda_1 \lambda_2} e_R^\mu + C_{s}^{\lambda_1 \lambda_2} e_{\lambda_1 \lambda_2}^\mu \right) \bra{N^{*}} F_\mu^{V} \ket{N} \\
&= 2M \frac{g_D g_{Z^\prime N}}{q^2 - m^2_{Z^\prime}} \bra{N^{*}} C_L^{\lambda_1 \lambda_2} F_-^{\lambda_1 \lambda_2} + C_R^{\lambda_1 \lambda_2} F_+^{\lambda_1 \lambda_2} + C_{s}^{\lambda_1 \lambda_2} F_{0}^{\lambda_1 \lambda_2} \ket{N}.
\end{split}
\end{equation}
Here $V_{\lambda_1 \lambda_2}^\mu \equiv \left[ \overline{u}_{p_3 \lambda_2} \gamma_\mu \left( g_V - g_A \gamma^5 \right) u_{p_1 \lambda_1} \right] \left( g^{\mu \nu} - q^\mu q^\nu / m^2_{Z^\prime}  \right)$ which interacts with the hadronic current. Their values can be expressed as: $V^\mu_{\lambda_1 \lambda_2} = \epsilon_\nu^{\lambda_1 \lambda_2} (g^{\mu \nu} - q^\mu q^\nu / M_{Z^\prime})$, where the 4-vector $\epsilon^\nu_{\lambda_1 \lambda_2}$ is defined as follows:

\begin{equation}
\label{eq:epsilon_for_vector}
\begin{split}
\epsilon^0_{\lambda_1 \lambda_2} &= \left( \alpha_{\lambda_1 \lambda_2} g_V \Delta_{+, \lambda_1 \cdot \lambda_2} - g_A \Delta_{-, \lambda_1 \cdot \lambda_2}\right) \sqrt{1 + \lambda_1 \cdot \lambda_2 \cos \delta}, \\
\epsilon^1_{\lambda_1 \lambda_2} &= \alpha_{\lambda_1 \lambda_2} \left( g_V \Delta_{-, \lambda_1 \cdot \lambda_2} - \beta_{\lambda_1 \lambda_2} g_A \Delta_{+, \lambda_1 \cdot \lambda_2} \right) \frac{\left| p_1 + \lambda_1 \cdot \lambda_2 p_2 \right|}{|q|} \sqrt{1 - \lambda_1 \cdot \lambda_2 \cos \delta}, \\
\epsilon^2_{\lambda_1 \lambda_2} &= i \left( \gamma_{\lambda_1 \lambda_2} g_V \Delta_{-, \lambda_1 \cdot \lambda_2} - g_A \Delta_{+, \lambda_1 \cdot \lambda_2}\right) \sqrt{1 - \lambda_1 \cdot \lambda_2 \cos \delta}, \\
\epsilon^3_{\lambda_1 \lambda_2} &= \gamma_{\lambda_1 \lambda_2} \left( g_V \Delta_{-, \lambda_1 \cdot \lambda_2} - \beta_{\lambda_1 \lambda_2} g_A \Delta_{+, \lambda_1 \cdot \lambda_2} \right) \frac{\left| p_1 - \lambda_1 \cdot \lambda_2 p_2 \right|}{|q|} \sqrt{1 + \lambda_1 \cdot \lambda_2 \cos \delta},
\end{split}
\end{equation}
where $\alpha_{\lambda_1 \lambda_2} = 1 - 2 \delta^{-}_{\lambda_1} \delta^{-}_{\lambda_2}$, $\beta_{\lambda_1 \lambda_2} = 1 - 2 \delta^{-}_{\lambda_1} \delta^{+}_{\lambda_2}$, and $\gamma_{\lambda_1 \lambda_2} = 1 - 2 \delta^{+}_{\lambda_1} \delta^{-}_{\lambda_2}$. Here, $\delta^i_j$ represents the Kronecker delta, $\lambda_i$ are the helicities of the incoming ($i = 1$) and outgoing ($i = 2$) DM fermions, $\delta$ is the angle between the DM fermions in the isobaric frame, and $\Delta_{\pm_1 \pm_2} \equiv \sqrt{E_1 E_3 \pm_1 m_\chi^2 \pm_2 p_1 p_3}$. It is important to note that all quantities in \Cref{eq:epsilon_for_vector} and their definitions are given in the isobaric frame and should not be confused with parameters in the Lab frame.

This vector, $V_{\lambda_1 \lambda_2}^\mu$, can be decomposed into components associated with different helicity states as $C_L^{\lambda_1 \lambda_2} e_L^\mu + C_R^{\lambda_1 \lambda_2} e_R^\mu + C_{s}^{\lambda_1 \lambda_2} e_{\lambda_1 \lambda_2}^\mu$, where $e^L,\, e^R,\, e_{\lambda_1\lambda_2}$ represent the basis vectors depending on the helicities,
\begin{equation}
\begin{split}
e_L^\mu &= \frac{1}{\sqrt{2}} \left( 0, 1, -i, 0 \right)\\
e_R^\mu &= \frac{1}{\sqrt{2}} \left( 0, -1, -i, 0 \right)\\
e_{\lambda_1 \lambda_2}^\mu &= \frac{1}{\sqrt{\left| \left(V_{\lambda_1 \lambda_2}^0 \right)^2 - \left(V_{\lambda_1 \lambda_2}^3 \right)^2 \right|}} \left( V_{\lambda_1 \lambda_2}^0, 0, 0, V_{\lambda_1 \lambda_2}^3 \right)
\end{split}
\end{equation}
and the values of the coefficients are
\begin{equation}
\begin{split}
C_L^{\lambda_1 \lambda_2} &= \frac{1}{\sqrt{2}} \left(V_{\lambda_1 \lambda_2}^1 + i V_{\lambda_1 \lambda_2}^2 \right)   \\
C_R^{\lambda_1 \lambda_2} &= - \frac{1}{\sqrt{2}} \left(V_{\lambda_1 \lambda_2}^1 - i V_{\lambda_1 \lambda_2}^2 \right)\\
C_s^{\lambda_1 \lambda_2} &= \mathrm{sign} \left( \left(V_{\lambda_1 \lambda_2}^0 \right)^2 - \left(V_{\lambda_1 \lambda_2}^3 \right)^2 \right) \sqrt{\left| \left(V_{\lambda_1 \lambda_2}^0 \right)^2 - \left(V_{\lambda_1 \lambda_2}^3 \right)^2 \right|}.
\end{split}
\end{equation}

For calculating the DM resonant  scattering differential cross section, we also consider helicity cross sections defined by:
\begin{equation}
\begin{split}
\sigma_L &= \frac{\pi W}{16 m_N p_1^2} \sum_{2j_z = 1,3} \left| f_{-2 j_z} \right|^2 \delta \left( W - M \right) \\
\sigma_R &= \frac{\pi W}{16 m_N p_1^2} \sum_{2j_z = 1,3} \left| f_{2 j_z} \right|^2 \delta \left( W - M \right) \\
\sigma_s^{\lambda_1 \lambda_2} &= \frac{\pi W}{16 m_N p_1^2} \sum_{k=\pm} \left| f_{0 k  }^{\lambda_1 \lambda_2} \right|^2 \delta \left( W - M \right)
\end{split}
\end{equation}
where the amplitudes are defined as:
\begin{equation}
\begin{split}
f_{\pm\left|2 j_z\right|} &= \bra{N, j_z \pm 1} F_\pm \ket{N^*, j_z }\,, \\
f_{0 \pm  }^{\lambda_1 \lambda_2} &= \bra{N, \pm \frac{1}{2}} F_{0}^{\lambda_1 \lambda_2} \ket{N^*, \pm \frac{1}{2} }\,.
\end{split}
\end{equation}
For the full list of amplitudes, we refer the reader to the vectorial helicity amplitudes in Table I in \cite{HoefkenZink:2025tns}.

\subsection{Computation of $V^S$ for scalar resonant scattering}
\label{sec:App_resonances_scalar_Vs}

In order to compute the resonant cross section through the scalar, we also need to compute the quantity $V^S_{\lambda_1 \lambda_2} \equiv \left[ \overline{u}_{p_3 \lambda_2}u_{p_1 \lambda_1} \right]$. The result is the following,

\begin{equation}
V^S_{\lambda_1 \lambda_2} = (1 - \delta_{-1, \lambda_1} \cdot \delta_{-1, \lambda_2}) \frac{\Delta^{S}_{\lambda_1 \lambda_2}}{A_S} \times \sqrt{1 + \lambda_1 \cdot \lambda_2 \cos \delta_{IB}},
\end{equation}
where $\Delta^{S}_{\lambda_1 \lambda_2} \equiv (E_1 + m_\chi)(E_3 + m_\chi) - \lambda_1 \cdot \lambda_2 |\vec{p}_1| |\vec{p}_3|$, $A_S \equiv \sqrt{2 (E_1 + m_\chi)(E_3 + m_\chi)}$ and $\delta_{IB}$ is the angle between the incoming and outgoing DM particle. Everything is considered in the already mentioned Isobaric frame.

\subsection{Scalar amplitudes}
\label{sec:App_resonances_scalar}

The amplitudes for the scalar case are defined,

\begin{equation}
f^S_{0^\pm} \equiv \bra{N^{*}, \pm \frac{1}{2}} F^S_{0^\pm} \ket{N, \pm \frac{1}{2}}.
\end{equation}
These amplitudes can be modelled using the FKR model~\cite{Feynman:1971wr}. We use the Hamiltonian with two harmonic oscillators of this model and add a Yukawa interaction between the three quarks in the system and the scalar: $g_{\Phi q} \overline{q} q \Phi$. $g_{\Phi q}$ is just the scalar coefficient and it is $g_u$ for the up quark and $g_d$ for the down quark. The $\Phi$ enters the computation, as with the vector case, as $e^{i q \cdot u_q}$, where $q$ is the 4-Momentum of the scalar and $u_q$ is the position of the quark. After following the same procedure performed by the authors, we find the following matrix operator element:
\begin{equation}
3 Q_a e^{- \lambda a_z}
\end{equation}
where $\lambda \equiv \sqrt{\frac{2}{\Omega}} \frac{m_N}{W} q_z$~\cite{Rein:1980wg} and $e_\alpha$ is the coupling element of the transition, as computed by FKR. The values of $Q_a$ depend on the representations of the baryons. For the scalar case, the non-vanishing values for an initial proton are:

\begin{equation}
\begin{split}
\bra{8_\alpha} Q_a \ket{10_S} &= \frac{\sqrt{2}}{3} (g_u - g_d)\\
\bra{8_\alpha} Q_a \ket{8_\alpha} &= \frac{1}{3} (g_u + 2 g_d)\\
\bra{8_\beta} Q_a \ket{8_\beta} &= g_u.
\end{split}
\end{equation}
Any other elements are zero. The only non-vanishing form factors found in this approach for an initial proton are the following,
\begin{center}
\renewcommand{\arraystretch}{2}
\begin{tabular}{|>{\centering\arraybackslash}p{2.5cm}|>{\centering\arraybackslash}p{5cm}|>{\centering\arraybackslash}p{5cm}|}
\hline
\textbf{Resonance} & \boldsymbol{$f^S_{0^+}$} & \boldsymbol{$f^S_{0^-}$} \\
\hline
$P_{11}$ (1440) & $- \frac{1}{6 \sqrt{3}} (2 g_u + g_d) S \lambda^2$ & $- \frac{1}{6 \sqrt{3}} (2 g_u + g_d) S \lambda^2$ \\
\hline
$D_{13}$ (1520) & $- \frac{1}{3 \sqrt{3}} (g_u - g_d) S \lambda$ & $- \frac{1}{3 \sqrt{3}} (g_u - g_d) S \lambda$ \\
\hline
$S_{11}$ (1535) & $\frac{1}{3 \sqrt{6}} (g_u - g_d) S \lambda$ & $- \frac{1}{3 \sqrt{6}} (g_u - g_d) S \lambda$ \\
\hline
$S_{31}$ (1620) & $\frac{1}{3 \sqrt{6}} (g_u - g_d) S \lambda$ & $- \frac{1}{3 \sqrt{6}} (g_u - g_d) S \lambda$ \\
\hline
$F_{15}$ (1680) & $\frac{1}{3 \sqrt{10}} (2 g_u + g_d) S \lambda^2$ & $\frac{1}{3 \sqrt{10}} (2 g_u + g_d) S \lambda^2$ \\
\hline
$D_{33}$ (1700) & $- \frac{1}{3 \sqrt{3}} (g_u - g_d) S \lambda$ & $- \frac{1}{3 \sqrt{3}} (g_u - g_d) S \lambda$ \\
\hline
$P_{11}$ (1710) & $\frac{1}{6 \sqrt{6}} (g_u - g_d) S \lambda^2$ & $\frac{1}{6 \sqrt{6}} (g_u - g_d) S \lambda^2$ \\
\hline
$P_{13}$ (1720) & $- \frac{1}{3 \sqrt{15}} (2 g_u + g_d) S \lambda^2$ & $\frac{1}{3 \sqrt{15}} (2 g_u + g_d) S \lambda^2$ \\
\hline
\end{tabular}
\end{center}
In order to simplify notation, we have omitted the $S$ superscript, so that: $g_u \equiv g^S_u$ and $g_d \equiv g^S_d$. For an initial neutron, we must replace $g_u$ with $g_d$ and vice versa. Here $S \equiv 3 G^S(q^2)$, where $G^S(q^2)$ is a scalar dipole, analogue to the vectorial and axial ones in Eq. $(3.12)$ in \cite{Rein:1980wg}:
\begin{equation}
G^S(q^2) \equiv G^S (0) \left( 1 - q^2 / m_S^2 \right)^{-2},
\end{equation}
where $m_S$ is an analogue to the vector and axial masses, but for the scalar, and $G^S (0)$ is the value of the form factor at $q^2 = 0$. Inspired in the values found by the authors in ~\cite{Alexandrou:2021ztx}, which were used for the pion in lattice QCD, we will set the scalar mass to $m_S = 1.22$ GeV and $G^S (0)$ to $1.17$.

\section{Amplitude and hadronic current}
\label{Ap:FF_details}
The different elements of the vector and axial  coefficients given in \Cref{eq:vector_axial_coefficients} are defined as
\begin{equation}
\label{eq:nucleon_currents}
\begin{split}
v_3^\nu &= \frac{1}{2} [\overline{u} \gamma^\nu u - \overline{d} \gamma^\nu d]\\
j_{A Q}^\nu &= \frac{2}{3} \sum_\alpha [\overline{q}_\alpha^U \gamma^\nu q_\alpha^U] - \frac{1}{3} \sum_\alpha [\overline{q}_\alpha^D \gamma^\nu q_\alpha^D]\\
v_s^\nu &= \sum_{q = s,c,b,t} \overline{q} \gamma^\nu q\\
a_3^\nu &= \frac{1}{2} [\overline{u} \gamma^\nu \gamma^5 u - \overline{d} \gamma^\nu \gamma^5 d]\\
a_0^\nu &= \frac{1}{2} [\overline{u} \gamma^\nu \gamma^5 u + \overline{d} \gamma^\nu \gamma^5 d]\\
a_s^\nu &= \sum_{q = s,c,b,t} \overline{q} \gamma^\nu \gamma^5 q.
\end{split}
\end{equation}
We use the usual conventions for the currents and include the axial current $a_0^\nu$.

\section{Details on the DM - Nucleon Scalar Interaction}
\label{sec:app_dm_nucleon}

To derive the expression for interactions mediated by a scalar, we start by considering the effective vector current, which is a linear combination of the Standard Model (SM) nucleon currents: \begin{equation} j_S^\mu = \frac{g_{\Phi u}}{2 m_u} \overline{u} \gamma^\mu u + \frac{g_{\Phi d}}{2 m_d} \overline{d} \gamma^\mu d + r^\mu \end{equation} where $r^\mu$ represents contributions independent of the $u$ and $d$ quarks. Contracting this expression with the sum of initial and final momenta yields: \begin{equation} (p_i^\mu + p_f^\mu) j_{S \mu} = g_{\Phi u} \overline{u} u + g_{\Phi d} \overline{d} d + r' \end{equation} with $r'$ again being independent of $u$ and $d$. This step makes use of the relations $\slashed{p}_a u_q(p_a) = m_q u_q(p_a)$ and $\overline{u}_q(p_a) \slashed{p}_a = \overline{u}_q(p_a) m_q$ for $a = i, f$.


Next, we contract the sum of initial and final momenta with the matrix element: \begin{equation} 
\begin{split} 
\Delta p_\mu  \bra{N(p_f)} j_{S}^\mu (0) \ket{N(p_i)} &=\Delta p_\mu \overline{u}N (p_f) \left[\gamma^\mu F_1^{S N} (Q^2) + \frac{i}{2 m_N} \sigma^{\mu \nu} q\nu F_2^{S N} (Q^2) \right] u_N(p_i)\\
&\simeq \overline{u}_N (p_f) \left[2 m_N F_1^{S N} (Q^2) - \frac{Q^2}{2 m_N} F_2^{S N} (Q^2) \right] u_N(p_i). 
\end{split} 
\end{equation} 
with $\Delta p_\mu\equiv(p_{i \mu} + p_{f \mu})$. This expression simplifies the scalar form factors in terms of the nucleon currents.


The scalar form factors $F_i^{SN}$ are related to the SM form factors as: 
\begin{equation}
F_i^{SN} \simeq \frac{3}{2} \left( \frac{g_{\Phi u}}{m_u} + \frac{g_{\Phi d}}{m_d} \right) F_i^N \mp \frac{1}{2} \left( \frac{g_{\Phi u}}{m_u} + 2 \frac{g_{\Phi d}}{m_d} \right) \left( F_i^p - F_i^n \right).
\end{equation} 
Finally, the differential cross section in the CM frame is calculated as: 
\begin{equation} 
\frac{d\sigma^N}{dz} = \frac{g_D^2 g_{\Phi N}^2}{8 \pi m_N^2 \left(E_1+E_2\right){}^2} \frac{\left(E_1^2 + m_{\chi }^2 - p_1^2 z\right) \left(p_1^2 (1+z) +2 m_N^2\right) \left(2 F_1^{\text{SN}} m_N^2-p_1^2 F_2^{\text{SN}} (1-z)\right)^2}{\left(2 p_1^2 (1-z) + m_{\Phi}^2\right)^2}
\end{equation} 
where $p_1 = \sqrt{E_1^2 - m_\chi^2}$, $z = \cos \theta$ and $E_2$ is the incoming nucleon total energy. Here, $g_{\Phi N}$ scales the nucleon couplings, and $g_{\Phi q}$ scales the quark couplings.

\section{DM - nucleon interaction with a scalar using Helm and Fermi - Symmetrized Woods - Saxon form factors}
\label{Ap:Helm_FF_computation}
For the sake of simplicity, we will assume the scalar is an isoscalar, so it interacts in the same way with protons and neutrons, so that we can model the interaction by using the Helm form factor~\cite{helm1956inelastic, lewin1996review, vietze2015nuclear}, which is regarded as the Fourier Transform of the density distribution of the nucleon. To compute the cross section, we will model the differential cross section following~\cite{fieguth2018discriminating}:

\begin{equation}
\begin{split}
\frac{d\sigma^N}{dQ^2} = &\frac{\sigma_0 E_\chi^2}{4 \mu_N (E_\chi^2 - m_\chi^2)} F_H^2 (Q^2),
\end{split}
\end{equation}
where $\sigma_0$ is the total cross section after considering an interaction with nucleons as point particles, $\mu_N = m_N m_\chi / (m_N + m_\chi)$ and $F_H^2 (Q^2)$ is the Helm form factor, equal to:

\begin{equation}
\label{eq:HelmFF}
\begin{split}
F_H^2 (Q^2) = \Bigg[ \frac{3 j_1 (q r_1)}{q r_1} \Bigg]^2 e^{- q^2 s^2}
\end{split}
\end{equation}
where $j_1 (x)$ is the spherical Bessel function of the first kind, $s$ is the nuclear skin thickness, $r_1$ is an effective nuclear radius. From the fitting done by~\cite{lewin1996review} by considering the Two-Parameter Fermi (Woods-Saxon) as the nuclear density model, we set:

\begin{equation}
\begin{split}
r_1 = \sqrt{c^2 + \frac{7}{3} \pi^2 a^2 - 5 s^2}
\end{split}
\end{equation}
such that $s \simeq 0.9$ fm, $a \simeq 0.523$ fm and $c \simeq 1.23 A^{1/3} - 0.60$ fm, for an atomic mass number $A$. The Two-Parameter Fermi model regards the nucleon density as a solid sphere surrounded by a thin shell~\cite{lewin1996review}.

$\sigma_0$ can be computed analytically, though its expression is long to be reproduced here. The form of the amplitude for this process is:

\begin{equation}
\begin{split}
i \mathcal{M} &= -i \frac{g_D g_{N\Phi}}{q^2 - m_\Phi^2} \big[\overline{u}_\chi (p_3) u_\chi (p_1) \big] \big[ \overline{u}_N (p_4) u_N (p_2) \big]. 
\end{split}
\end{equation}
And then the final result is obtained by computing:
\begin{equation}
\label{eq:sigma_0}
\begin{split}
\sigma_0 = \frac{g_D^2 g_{N\Phi}^2}{8 \pi p_\chi^2} \int_0^{x_m} dx \frac{(x + 2 m_N^2) (x + 2 m_\chi^2)}{(x + m_N^2) (2x + m_\Phi^2)^2}
\end{split}
\end{equation}
such that $x_m = m_N E_\chi - m_N (m_N^2 E_\chi + m_\chi^2 (E_\chi + 2 m_N)) / (2 m_N E_\chi + m_N^2 + m_\chi^2)$ and $x$ was obtained by a change of variable: $x \equiv m_N (E_\chi - E_3)$. $E_\chi$ and $p_\chi$ are the incoming DM particle energy and $3D$-momentum respectively and $E_3$ is the energy of the outgoing DM particle. Eq.~\eqref{eq:sigma_0} can be computed explicitly and used in the numerical computation of the total cross section.

One variation of this model consists on using the Fermi - Symmetrized Woods - Saxon form factors. We would need to replace the Helm one in \Cref{eq:HelmFF} by the following one~\cite{grypeos1991cosh, kamp2023dipole}:

\begin{equation}
\begin{split}
F^{FS-WS} (Q) = \frac{3 \pi a}{r_0^2 + \pi^2 a^2} \frac{a \pi \coth \left(\pi Q a \right) \sin \left(Q r_0 \right) - r_0 \cos \left(Q r_0 \right)}{Q r_0 \sinh \left(\pi Q a \right)},
\end{split}
\end{equation}
where $a$ is the same as the one defined for the Helm form factor and $r_0 \equiv 1.03 \times A^{1/3}$ fm. This is the one we will use for our computations, instead of that of Helm.




\bibliographystyle{JHEP}
\bibliography{lib}

\end{document}